\documentclass[journal]{IEEEtran}
\usepackage{cite}
\usepackage{amsmath}
\usepackage{amssymb}
\usepackage{amsfonts}
\usepackage{stfloats}
\usepackage{graphicx}
\usepackage{algorithm}
\usepackage{algorithmic}
\usepackage{epsfig}
\usepackage{subfigure}
\usepackage[draft,defaultcolor=blue]{changes}  
\usepackage{xcolor}
\colorlet{added}{blue} 

\usepackage{pifont}
\usepackage{psfrag}
\usepackage{times}
\usepackage{booktabs}
\usepackage[T1]{fontenc}
\usepackage{moreverb}
\usepackage{color}
\usepackage{setspace}
\usepackage{mathrsfs}
\usepackage{array}
\usepackage{txfonts}
\usepackage{makecell}
\usepackage{afterpage}

\begin{document}

\title{Waveform Design for OTFS Assisted Simultaneous Acoustic Information and Power Transfer}

\author{Jinheng~Kang, Yizhe~Zhao,~\IEEEmembership{Member,~IEEE}, Kun~Yang,~\IEEEmembership{Fellow,~IEEE}, Jun~Liu,~\IEEEmembership{Member,~IEEE}
\thanks{Jinheng Kang and Yizhe Zhao are with the School of Information and Communication Engineering, University of Electronic Science and Technology of China, Chengdu 611731, China,  email: jhkang@std.uestc.edu.cn, yzzhao@uestc.edu.cn.}

\thanks{Kun Yang is with State Key Laboratory of Novel Software Technology, Nanjing
	University, Nanjing 210008, China, and School of Intelligent Software and
	Engineering, Nanjing University (Suzhou Campus), Suzhou 215163, China,
	e-mail: kunyang@nju.edu.cn.}
\thanks{Jun Liu is with the School of Electronic and Information Engineering, Beihang
	University, Beijing 100191, China, e-mail: liujun2019@buaa.edu.cn.}
}

\maketitle

\begin{abstract}
	Simultaneous acoustic information and power transfer (SAIPT) is a promising technique for supporting self-sustainable Internet of Underwater Things (IoUT) networks through concurrent data transmission and energy supplement. However, existing OFDM-based SAIPT studies are vulnerable to severe multipath propagation and Doppler effects in dynamic underwater acoustic channels. To address this issue, this paper proposes an orthogonal time frequency space (OTFS)-based SAIPT waveform design for dynamic underwater acoustic channels. The acoustic information transfer (AIT) and acoustic power transfer (APT) symbols are jointly designed, while the transducer conversion efficiencies and nonlinear rectifier characteristics are incorporated into the system model. Based on the derived achievable data rate and DC output expressions, a waveform optimization problem is formulated to maximize the harvested DC output under transmit power and minimum data-rate constraints. To solve the resulting non-convex problem, a successive convex approximation (SCA)-based algorithm is developed. Simulation results show that the proposed OTFS-based design outperforms the OFDM-based scheme in terms of DC output  in the dynamic transmission scenarios. The effects of key system parameters are also analyzed, confirming the effectiveness of the proposed design in improving acoustic energy transfer efficiency.
\end{abstract}

\begin{IEEEkeywords}
Orthogonal-time-frequency-space (OTFS), simultaneous acoustic information and power transfer (SAIPT), waveform design

\end{IEEEkeywords}

\section{Introduction}
\subsection{Background}
	The rapid development of marine exploration and underwater intelligent systems has significantly increased the demand for reliable underwater sensing, communication, and networking technologies. With the expansion of applications such as marine environmental monitoring, offshore infrastructure inspection, underwater surveillance, and resource exploration, underwater networks are expected to support long-term, large-scale, and autonomous operation in complex marine environments~\cite{linUnderwaterPollutionTracking2023,liBinocularUnderwaterMeasurement2024}. In this context, the Internet of Underwater Things (IoUT) has emerged as an important networking paradigm for connecting underwater sensors, autonomous underwater vehicles (AUVs), and other intelligent devices in maritime environments. IoUT supports a wide range of applications, such as underwater exploration, environmental monitoring, positioning and navigation, disaster prediction, and maritime security~\cite{jahanbakhtInternetUnderwaterThings2021}.
	
	Despite its promising application prospects, the practical deployment of IoUT still faces two fundamental challenges. On the one hand, underwater nodes are usually powered by batteries and deployed in deep-sea or hazardous environments, making battery replacement and maintenance difficult, costly, and sometimes impractical~\cite{sendraUnderwaterAcousticModems2016}. On the other hand, underwater acoustic channels are severely affected by multipath propagation and Doppler effects, resulting in highly time-varying channel conditions and posing significant challenges to reliable information transmission, especially in high-mobility scenarios involving AUVs or drifting nodes~\cite{stojanovic2009underwater,liuDoublySelectiveUnderwater2012}.
	
	To improve the energy sustainability of underwater networks, simultaneous acoustic information and power transfer (SAIPT) has been proposed as a promising technique for supporting both information transmission and energy delivery through acoustic waves. By using acoustic signals as a unified carrier, SAIPT enables an underwater transmitter to convey data while simultaneously replenishing the energy of underwater sensor nodes, thereby reducing the dependence on battery replacement and extending the lifetime of IoUT networks~\cite{luoSimultaneousWirelessPower2024,yizhiWirelessInformationPower2024}. 
	
	However, when conventional OFDM-based waveform design is applied to SAIPT systems, its performance can be significantly degraded in dynamic underwater acoustic channels. In particular, Doppler effects destroy the orthogonality among subcarriers and lead to severe inter-carrier interference, especially when the transmitter or receiver is mounted on a moving platform such as an AUV~\cite{tuMultipleResamplingReceiver2013}. Therefore, a more robust waveform design is needed to improve both information transmission and acoustic energy transfer performance in high-mobility underwater scenarios.
	
	Orthogonal time frequency space (OTFS) modulation has recently emerged as an effective technique for high-mobility communication systems. Unlike OFDM, which represents signals in the time-frequency domain, OTFS maps information symbols in the delay-Doppler domain, where the time-varying channel can be represented in a more stable and compact form. This property enables OTFS to exploit the delay-Doppler domain sparsity of underwater acoustic channels and provides improved robustness against Doppler shifts and multipath propagation~\cite{yang2026underwater,zhangOTFSBasedUnderwater2026}.

\subsection{Related Works}
	Underwater acoustic communication forms the foundation of IoUT networks, and extensive studies have been conducted to improve communication reliability in challenging underwater environments. For example, Zhu \textit{et al.} proposed an orthogonal chirp division multiplexing (OCDM)-based multi-carrier scheme to enhance robustness against multipath fading in underwater acoustic channels ~\cite{zhuAntiMultipathOCDM2020}. Sun \textit{et al.}~\cite{sunSymbolBasedPassbandDoppler2020} developed a symbol-by-symbol passband Doppler compensation method for DSSS underwater acoustic signals. In addition, multicarrier modulation has been widely investigated in underwater acoustic systems. Esmaiel \textit{et al.}~\cite{esmaiel2022energy} proposed a TDS-OFDM-based multicarrier system combined with NOMA to improve spectral efficiency, while Liu \textit{et al.}~\cite{liu2025densenet} developed a DenseNet-based channel estimation method for underwater acoustic OFDM systems. Recently, OTFS has also been introduced into underwater acoustic communication to cope with time-varying multipath channels and Doppler effects. Wang \textit{et al.}~\cite{wangIterativeLMMSESIC2024} studied Doppler-squint-aware UWA-OTFS systems and proposed an LMMSE-SIC detection scheme for reliable data recovery in double-dispersive underwater acoustic channels. Jing \textit{et al.}~\cite{jingDirectAdaptiveTurbo2025} investigated OTFS-based mobile UWA communication and developed an adaptive turbo equalization scheme to improve data detection performance under high-mobility conditions. Yang \textit{et al.}~\cite{yangInterDopplerOCDFE2026} proposed an iterative decision feedback equalizer with optimal combining for UWA-OTFS systems, achieving a tradeoff between bit-error-rate performance and computational complexity.
	
	Despite the advances in underwater acoustic communication, the limited battery capacity of underwater devices remains a critical challenge for long-term and sustainable IoUT operations. To address this issue, simultaneous wireless information and power transfer (SWIPT) has attracted increasing attention as a promising solution for enabling concurrent energy delivery and data transmission. Extensive studies have investigated RF-based SWIPT in terrestrial wireless networks, where joint waveform, beamforming, and resource allocation designs have been developed to improve energy harvesting efficiency and communication performance. For instance, Zhang et al. investigated fluid-antenna-assisted integrated data and energy transfer (IDET) systems and jointly optimized port selection and beamforming design to improve the energy–rate trade-off~\cite{zhang2024joint}. Furthermore, they considered practical switching delay and energy consumption of port selection and developed energy-efficient strategies from both short-term and long-term perspectives~\cite{zhangEnergyEfficientPortSelection2025a}. Bian \textit{et al.}\cite{bianQoSAwareEnergyStorage2023} studied RIS-assisted SWIPT networks and optimized RIS coefficients and task allocation for energy storage enhancement. Lin \textit{et al.}\cite{lin2025fluid} investigated a fluid-antenna-assisted IDET system and optimized port selection strategies to improve the performance trade-off between data transmission and energy harvesting. Zhang \textit{et al.}\cite{zhang2025performance} investigated a pinching-antenna-enabled SWIPT system and optimized antenna deployment and transmission protocols to balance information transmission and energy harvesting.
		
	However, RF-based SWIPT is difficult to apply in underwater environments due to the severe attenuation of electromagnetic waves in seawater. Therefore, simultaneous acoustic information and power transfer (SAIPT) has emerged as a more suitable solution for underwater applications. Xing \textit{et al.}~\cite{xing2023performance} analyzed the performance of a SAIPT-enabled system and characterized its bit-error-rate and outage behavior. Esmaiel \textit{et al.}~\cite{esmaiel2020wireless} investigated a time-reversed NOMA scheme for underwater SAIPT. Omeke \textit{et al.}~\cite{omekeSustainableInternetUnderwater2024a} proposed a reinforcement learning-based SAIPT scheme using an AUV to jointly improve data throughput and wireless power transfer. Deepa \textit{et al.}~\cite{deepaEnhancedSWIPTCooperative2025} studied cooperative relaying for NOMA-based underwater acoustic sensor networks, while Melki \textit{et al.}~\cite{melkiAUVTrajectoryLearning2025} developed a deep reinforcement learning-based SAIPT framework considering data freshness, energy harvesting, and fairness.
	
\subsection{Motivations and Contributions}	
	Despite the recent progress in SAIPT, existing studies still face the following limitations:
	\begin{itemize}
		\item \textit{Simplified device modeling:} Practical device characteristics are often simplified in current SAIPT models. For instance, \cite{xing2023performance} assumes a frequency-independent transducer conversion efficiency, which cannot capture the practical frequency-selective characteristics of acoustic transducers. In addition, \cite{esmaiel2020wireless} investigates a SAIPT scheme where the diode nonlinearity is initially modeled using a Taylor series expansion, but the harvested DC output is ultimately approximated by only the second-order term, thereby neglecting higher-order nonlinear effects. These simplifications may lead to inaccurate characterization of acoustic energy harvesting performance and limit the effectiveness of waveform optimization in practical underwater acoustic environments.
		
		\item \textit{Insufficient waveform design for dynamic underwater acoustic channels:} The waveform design for dynamic underwater acoustic channels remains insufficiently explored in existing SAIPT studies. When conventional OFDM-based waveform design is applied to SAIPT systems, Doppler effects may destroy the orthogonality among subcarriers and introduce severe inter-carrier interference, especially in high-mobility scenarios involving AUVs or drifting nodes~\cite{tuMultipleResamplingReceiver2013,liuDoublySelectiveUnderwater2012}. 
	\end{itemize}
		
	To address these issues, this paper proposes an OTFS-based SAIPT waveform design for 
	dynamic underwater acoustic channels. The proposed framework jointly considers the 
	delay-Doppler domain robustness of OTFS, the frequency-dependent conversion 
	characteristics of acoustic transducers, and the nonlinear energy conversion behavior of 
	rectifiers. The main contributions of this paper are summarized as follows:
	\begin{itemize}
		\item To the best of our knowledge, this is the first work to introduce OTFS into underwater SAIPT waveform design. A practical OTFS-based SAIPT system model is established, where information and power transfer signals are jointly designed in the delay-Doppler domain to improve robustness against time-varying multipath propagation and Doppler effects. In addition, the frequency-dependent transducer conversion efficiencies and nonlinear rectifier characteristics are incorporated to accurately characterize the harvested DC output.
		
		\item Based on the derived achievable-rate and DC-output expressions, a non-convex 
		waveform optimization problem is formulated to maximize the harvested DC output 
		under the transmit power constraint and the minimum data-rate requirement. To solve 
		this problem, an SCA-based iterative algorithm is developed, where the non-convex 
		objective function and rate constraint are transformed into tractable convex 
		approximations.
		
		\item Simulation results demonstrate that the proposed OTFS-based SAIPT waveform design achieves a higher DC output than the OFDM-based benchmark. Moreover, by incorporating the transducer frequency response and nonlinear rectification behavior, the proposed design can effectively exploit the high PAPR of multicarrier waveforms, thereby significantly improving the acoustic power transfer efficiency.		
	\end{itemize}
	
	\begin{figure*}[!t]
	\centering
	\includegraphics[width=0.98\textwidth]{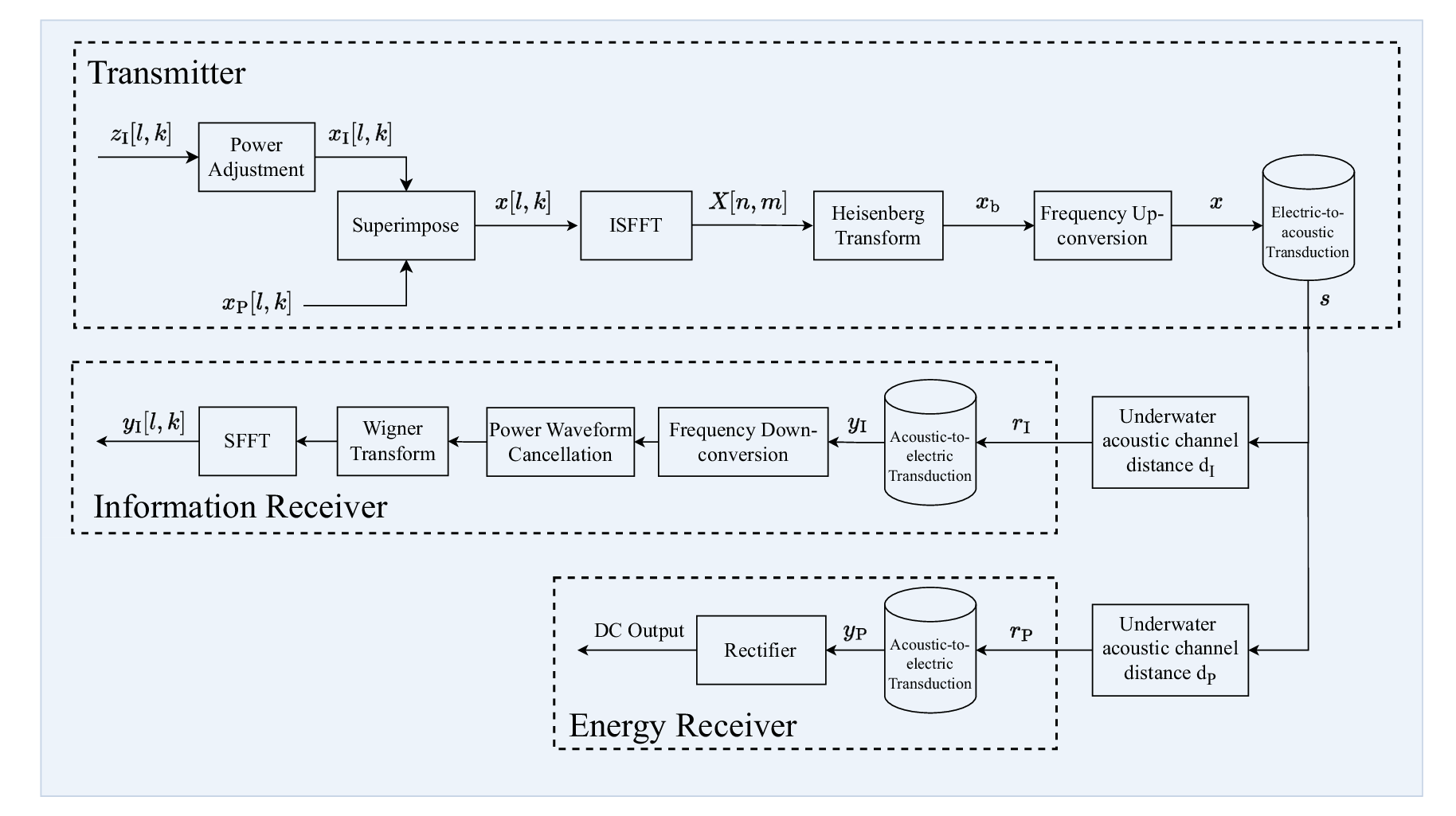}
	\caption{SAIPT system model.}
	\label{fig:system_flowchart}
	\end{figure*}
	
	The rest of this paper is organized as follows: The model of OTFS-based SAIPT system is introduced in Section \ref{sec:SYSTEM MODEL}. The formulation and solution of the waveform optimization problem are presented in Section~\ref{sec:WAVEFORM DESIGN}. Numerical results are illustrated in Section \ref{NUMERICAL RESULTS}. Finally, our paper is concluded in Section \ref{sec:CONCLUSION}.
	
	\textit{Notations:} In this paper, boldface upper letters, boldface lower letters, and lower letters denote matrices, vectors, and scalars, respectively. 
	$\mathbb{C}^{m \times n}$ and $\mathbb{R}^{m \times n}$ represent the set of complex matrices and real matrices with the dimension of $m \times n$, respectively. 
	A complex vector of dimension $m$ is denoted in $\mathbb{C}^{m}$, while a real vector of dimension $m$ is denoted in $\mathbb{R}^{m}$. 
	$|x|$ represents the amplitude of a complex number $x$. 
	$\|\mathbf{x}\|$  denotes the 2-norm of a vector $\mathbf{x}$. 
	The notation $(\cdot)^{T}$ and $(\cdot)^{H}$ refer to the transpose and the conjugate transpose of a vector or matrix. $\mathbb{E}$ represents the expectation operator, while $\mathcal{P}\{\cdot\}$ refers to the DC component of a signal. $\Re\{\cdot\}$ and $\Im\{\cdot\}$ denote the real and imaginary parts of a complex vector or matrix, respectively.

\section{SYSTEM MODEL} \label{sec:SYSTEM MODEL}
\subsection{SAIPT Model} \label{sec:SAIPT MODEL}

	An OTFS-based SAIPT system is illustrated in Fig.~\ref{fig:system_flowchart}, 
	which consists of a SAIPT transmitter, an information receiver, and an energy receiver. Let $\mathbf{X}^{\mathrm{DD}}_{\mathrm{I}}$ denote the $M\times N$ information 
	signal matrix in the delay-Doppler (DD) domain, whose element is given by
	$
	x_{\mathrm{I}}[l,k]=z_{\mathrm{I}}[l,k]w_D[l,k],
	$
	where $l=0,1,\ldots,M-1$ and $k=0,1,\ldots,N-1$ denote the delay and Doppler 
	indices, respectively. Here, $M$ and $N$ denote the numbers of subcarriers and 
	time slots, respectively, $z_{\mathrm{I}}[l,k]\sim\mathcal{CN}(0,1)$ is the 
	random information symbol, and $w_D[l,k]$ denotes the corresponding power-control coefficient. Let $\mathbf{X}^{\mathrm{DD}}_{\mathrm{P}}$ denote the deterministic $M\times N$ power-transfer signal matrix with elements $x_{\mathrm{P}}[l,k]$. Unlike random information symbols, deterministic power-transfer symbols can be deliberately designed to generate high-PAPR waveforms and enhance the harvested DC output through nonlinear rectification. The superimposed DD-domain signal can then be expressed as
	$
		\mathbf{X}^{\mathrm{DD}}=\mathbf{X}^{\mathrm{DD}}_{\mathrm{I}}+\mathbf{X}^{\mathrm{DD}}_{\mathrm{P}},
	$
	where $\mathbf{X}^{\mathrm{DD}}=\{x[l,k]\}\in\mathbb{C}^{M\times N}$.
	The superimposed DD-domain signal $\mathbf{X}^{\mathrm{DD}}$ is then transformed into the TF domain by the inverse symplectic finite Fourier transform (ISFFT). In the TF domain, the frequency of the $m$-th
	subcarrier is given by
	\begin{equation}
		f_m = f_r + \left(m-\frac{M-1}{2}\right)\Delta f,
		\quad m=0,1,\ldots,M-1,
		\label{eq:subcarrier_frequency}
	\end{equation}
	where $f_r$ denotes the center frequency of the transducer, and $\Delta f$
	denotes the subcarrier frequency spacing. The obtained TF-domain signal matrix $\mathbf{X}^{\mathrm{TF}}$, which is denoted by $\mathbf{X}$ for notational simplicity in the sequel, is then converted into a continuous-time baseband waveform $x_{\mathrm{b}}$ through the Heisenberg transform. Subsequently, frequency up-conversion is performed to shift the 
	baseband waveform to the desired carrier frequency. The resulting electrical 
	signal $x$ is fed into the electro-acoustic transducer and converted into the acoustic 
	SAIPT signal $s$, which is then transmitted through the underwater acoustic channel.
	
	After propagation, the transmitted acoustic signal arrives at both the information receiver and the energy receiver. The propagation distances from the transmitter to 
	the information receiver and the energy receiver are denoted by $d_{\mathrm{I}}$ and 
	$d_{\mathrm{P}}$, respectively. At the information receiver, the received acoustic 
	signal $r_\mathrm{I}$ is first converted into the electrical signal $y_\mathrm{I}$ by the acoustic-electric transducer and then down-converted to the baseband. Subsequently, the power-transfer waveform is cancelled before information demodulation. The remaining signal is then processed by the Wigner transform and the symplectic finite Fourier transform (SFFT) to recover the information symbols $y_\mathrm{I}[l,k]$ in the DD 
	domain.
	
	At the energy receiver, the received acoustic signal $r_\mathrm{P}$ is also converted into the electrical signal $y_\mathrm{P}$ by the acoustic-electric transducer and then fed into the rectifier to obtain the DC output.
	
\subsection{Underwater Channel Model}
\subsubsection{Channel Propagation Model}
	\label{subsubsec:underwater_channel}
	
	The channel coefficient $h(\tau,\nu)$ in DD-domain is modeled as \cite{naikoti2022performance}:
	\begin{equation}
		h(\tau,\nu)
		=
		\sum_{p=1}^{N_{\mathrm{path}}} A_p \delta(\tau-\tau_p)\delta(\nu-\nu_p),
		\label{eq:delay_doppler_channel_continuous}
	\end{equation}
	where $N_{\mathrm{path}}$ denotes the number of propagation paths, and $A_p$ represents the 
	complex gain of the $p$-th path. Moreover, $\tau_p$ and $\nu_p$ denote the delay 
	and Doppler shift of the $p$-th path, respectively. In this work, the multipath 
	parameters $\{A_p,\tau_p\}_{p=1}^{P}$ are obtained from the Bellhop-based 
	underwater acoustic propagation model~\cite{porter2011bellhop}. The Doppler shift 
	of the $p$-th path is modeled as $\nu_p=a_p f_c$, where $f_c$ is the carrier 
	frequency and $a_p$ denotes the Doppler scaling factor. Considering an AUV with 
	the maximum velocity $v_{\max}$, $a_p$ is assumed to follow
	$
		a_p \sim 
		\mathcal{U}\left(-\frac{v_{\max}}{c},\frac{v_{\max}}{c}\right),
		\label{eq:doppler_factor_distribution}
	$ where $c$ denotes the underwater sound speed.
	The delay and Doppler resolutions of the DD domain are given by
	$
		\Delta \tau = \frac{1}{M\Delta f},
		\Delta \nu = \frac{1}{NT}.
		\label{eq:dd_domain_resolution}
	$
	Here, $M$, $\Delta f$, $N$, and $T$ are defined in the TF domain and are consistent 
	with their OFDM counterparts. Specifically, $M$ and $N$ denote the numbers of 
	subcarriers and time slots, respectively, while $\Delta f$ and $T$ represent the 
	subcarrier spacing and symbol duration, respectively.
	
	The corresponding TF-domain channel coefficient $H[m,n]$ can be obtained from 
	the DD-domain channel coefficient $h[l,k]$ as
	\begin{equation}
		H[m,n]
		=
		\sum_{l=0}^{M-1}
		\sum_{k=0}^{N-1}
		h[l,k]
		e^{j2\pi\left(\frac{nk}{N}-\frac{ml}{M}\right)},
		\label{eq:dd_to_tf_channel}
	\end{equation}
	where $m=0,1,\ldots,M-1$ and $n=0,1,\ldots,N-1$ denote the subcarrier index and 
	time-slot index in the TF domain, respectively. Here, $h[l,k]$ denotes the 
	discrete DD-domain channel coefficient at the $l$-th delay bin and the $k$-th 
	Doppler bin. The coefficient $h[l,k]$ is obtained by mapping the continuous 
	DD-domain channel response $h(\tau,\nu)$ onto the discrete DD grid according to 
	the delay and Doppler resolutions $\Delta\tau$ and $\Delta\nu$.
	
\subsubsection{Underwater Noise}
	Underwater acoustic noise consists of four main components: turbulence, shipping, wind, and thermal noise. 
	The overall power spectral density (p.s.d.) is modeled as the sum of these components:
	
	\begin{equation}
	N(f)=N_{t}(f)+N_{s}(f)+N_{w}(f)+N_{th}(f),
	\end{equation}
	where $f$ denotes the signal frequency. Detailed expressions of each noise component can be found in \cite{basagniAdvancesUnderwaterAcoustic2013}.
	
	The noise power over the system bandwidth $B$ is then obtained by integrating the p.s.d. as \cite{wangOptimalPowerAllocation2020a}
		\begin{align}
			P_{n}&=10\log_{10}\left(\int_{f_0}^{f_0+B}N(f)df\right),\\
			\sigma_{n}^{2}&=10^{\frac{P_{n}-10\log_{10}\phi-171.5}{10}},
		\end{align}
		where $f_{0}$ denotes the minimum operational frequency, $\phi$ represents the overall electro-acoustic efficiency, and $\sigma_n^2$ denotes the underwater noise power in watts.

\subsection{Transducer Efficiency Modeling and Fitting} 
	Let $\eta_{\mathrm{ea}}(f)$ denote the electro-acoustic conversion efficiency of 
	the transmitting transducer, which is defined as
	\begin{equation}
		\eta_{\mathrm{ea}}(f)
		=
		\frac{P_{\mathrm{ac},1}(f)}{P_{\mathrm{elec}}(f)},
		\label{eq:eta_ea_definition}
	\end{equation}
	where $P_{\mathrm{elec}}(f)$ is the input electrical power and 
	$P_{\mathrm{ac},1}(f)$ is the generated acoustic power evaluated at $1~\mathrm{m}$ 
	from the transducer. Based on the transmitting voltage response $TVR(f)$, the 
	directivity factor $D_f$, and the complex impedance $Z(f)$, the electro-acoustic 
	conversion efficiency can be expressed as
	\begin{equation}
		\eta_{\mathrm{ea}}(f)
		=
		1.194\times10^{-13}
		\frac{D_f}{f^2\Re\{1/Z(f)\}}
		10^{TVR(f)/10}.
		\label{eq:eta_ea_tvr}
	\end{equation}
	
	Similarly, let $\eta_{\mathrm{ae}}(f)$ denote the acousto-electric conversion 
	efficiency of the receiving transducer, which is defined as
	\begin{equation}
		\eta_{\mathrm{ae}}(f)
		=
		\frac{P_{\mathrm{harv}}(f)}{P_{\mathrm{ac}}(f)},
		\label{eq:eta_ae_definition}
	\end{equation}
	where $P_{\mathrm{ac}}(f)$ is the received acoustic power and 
	$P_{\mathrm{harv}}(f)$ is the harvested alternating-current power on the load. 
	For a reciprocal transducer, the receiving voltage sensitivity can be related to 
	the transmitting voltage response and impedance. Accordingly, the acousto-electric 
	conversion efficiency is given by
	\begin{equation}
		\eta_{\mathrm{ae}}(f)
		=
		8.32\times10^{-18}
		\frac{|Z(f)|^2}{D_f\Re\{Z(f)\}}
		10^{TVR(f)/10}.
		\label{eq:eta_ae_tvr}
	\end{equation}
		
	Since the measured transducer parameters are available only at limited sampling 
	frequency points, curve fitting is required to obtain continuous efficiency 
	functions over the entire operating band for subsequent system modeling and 
	waveform optimization. In this work, the electro-acoustic conversion efficiency 
	$\eta_{\mathrm{ea}}(f)$ and the acousto-electric conversion efficiency 
	$\eta_{\mathrm{ae}}(f)$ are fitted using the MATLAB Curve Fitting 
	Toolbox. The fitting results are summarized in 
	Tables~\ref{tab:eta_fitting}.
	
	\begin{table*}[!t]
	\centering
	\caption{Curve fitting results for transducer efficiencies}
	\label{tab:eta_fitting}
	\small
	\setlength{\tabcolsep}{3pt} 
	\renewcommand{\arraystretch}{0.95}
	
	\textbf{(a) Curve fitting results for $\eta_{\mathrm{ea}}$}
	
	\vspace{0.5em}
	
	\begin{tabular}{lccccc}
		\toprule
		\textbf{Fitting method} &
		\textbf{$R^2$} &
		\makecell{\textbf{Sum of}\\\textbf{Squared Errors}} &
		\makecell{\textbf{Adjusted}\\\textbf{$R^2$}} &
		\makecell{\textbf{Root Mean}\\\textbf{Squared Error}} &
		\makecell{\textbf{Number of}\\\textbf{coefficients}} \\
		\midrule
		Sinusoidal sum & 0.97089  & $4.8108\times10^{-6}$ & 0.96220  & 0.00024996 & 24 \\
		Fourier        & 0.96731  & $5.4037\times10^{-6}$ & 0.96061  & 0.00025516 & 18 \\
		Polynomial     & 0.95021  & $8.2292\times10^{-6}$ & 0.94529  & 0.00030072 & 10 \\
		Power          & 0.50664  & $8.1545\times10^{-5}$ & 0.49657  & 0.00091219 & 3  \\
		Logarithmic    & 0.48635  & $8.4898\times10^{-5}$ & 0.48116  & 0.00092605 & 2  \\
		Exponential    & 0.46443  & $8.8522\times10^{-5}$ & 0.44787  & 0.00095530 & 4  \\
		Sigmoidal      & 0.46232  & $8.8871\times10^{-5}$ & 0.44569  & 0.00095718 & 4  \\
		Rational       & 0.43488  & $9.3405\times10^{-5}$ & 0.37209  & 0.00101870 & 11 \\
		\bottomrule
	\end{tabular}
	
	\vspace{1.5em}
	
	\textbf{(b) Curve fitting results for $\eta_{\mathrm{ae}}$}
	
	\vspace{0.5em}
	
	\begin{tabular}{lccccc}
		\toprule
		\textbf{Fitting method} &
		\textbf{$R^2$} &
		\makecell{\textbf{Sum of}\\\textbf{Squared Errors}} &
		\makecell{\textbf{Adjusted}\\\textbf{$R^2$}} &
		\makecell{\textbf{Root Mean}\\\textbf{Squared Error}} &
		\makecell{\textbf{Number of}\\\textbf{coefficients}} \\
		\midrule
		Fourier        & 0.99650  & 0.00067231 & 0.99578  & 0.0028461 & 18 \\
		Gaussian       & 0.99268  & 0.0014052  & 0.99049  & 0.0042719 & 24 \\
		Polynomial     & 0.98646  & 0.0025984  & 0.98512  & 0.0053436 & 10 \\
		Exponential    & 0.97934  & 0.0039651  & 0.97870  & 0.0063935 & 4  \\
		Rational       & 0.97540  & 0.0047214  & 0.97266  & 0.0072430 & 11 \\
		Sinusoidal sum & 0.97425  & 0.0049407  & 0.96656  & 0.0080103 & 24 \\
		Power          & 0.90889  & 0.017484   & 0.90703  & 0.013357  & 3  \\
		Sigmoidal      & 0.73953  & 0.049983   & 0.73147  & 0.022700  & 4  \\
		Logarithmic    & 0.62657  & 0.071659   & 0.62280  & 0.026904  & 2  \\
		\bottomrule
	\end{tabular}
	\end{table*}
	
	For $\eta_{\mathrm{ea}}(f)$, the sinusoidal sum model achieves the best fitting 
	performance among the considered fitting methods, accurately capturing the 
	frequency-dependent resonant characteristics of the electro-acoustic conversion 
	process. Accordingly, the fitted electro-acoustic conversion efficiency is modeled as
	\begin{equation}
		\eta_{\mathrm{ea,fit}}(f)
		=
		\sum_{i=1}^{8} a_{\mathrm{ea},i}
		\sin\left(b_{\mathrm{ea},i}f+c_{\mathrm{ea},i}\right),
		\label{eq:eta_ea_fitting}
	\end{equation}
	where $a_{\mathrm{ea},i}$, $b_{\mathrm{ea},i}$, and $c_{\mathrm{ea},i}$ denote 
	the amplitude coefficient, frequency scaling factor, and phase offset of the 
	$i$-th sinusoidal basis function, respectively.
	
\begin{figure*}[htbp]
		\centering
		{\subfigure[]{\includegraphics[width=0.4\linewidth]{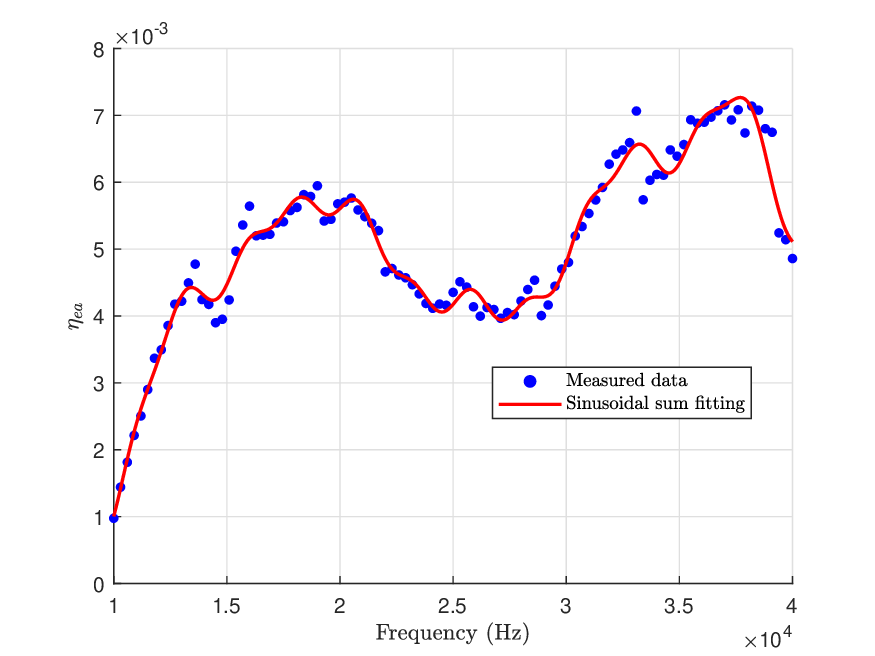}}
			\hfil
			\subfigure[]{\includegraphics[width=0.4\linewidth]{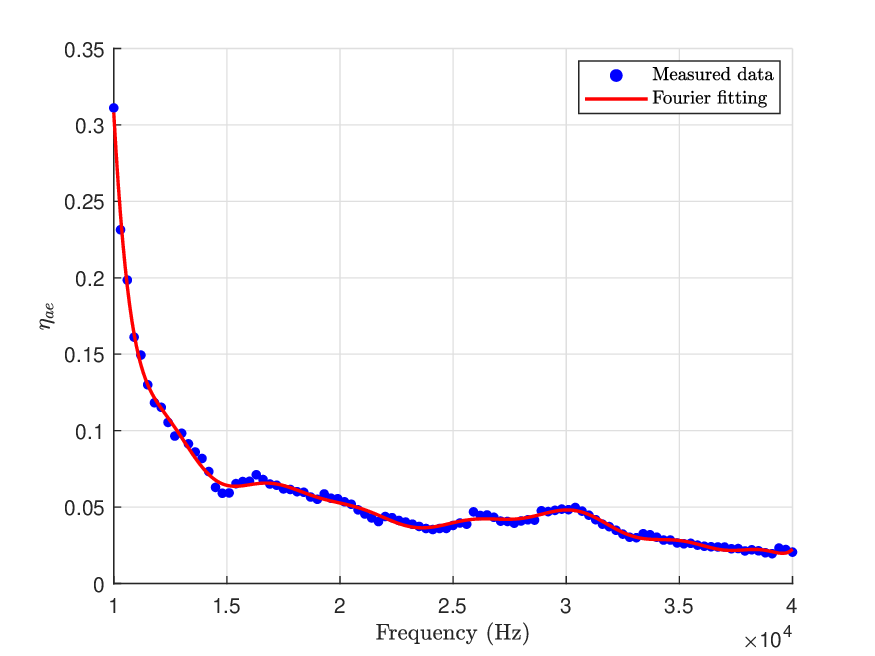}}}
		\setlength{\abovecaptionskip}{0pt}
		\setlength{\belowcaptionskip}{0pt}
		\caption{Fitting curves of transducer conversion efficiencies: 
			(a) electro-acoustic conversion efficiency $\eta_{\mathrm{ea}}(f)$; 
			(b) acousto-electric conversion efficiency $\eta_{\mathrm{ae}}(f)$.}
		\label{fig:eta_fitting}
	\end{figure*}

	For $\eta_{\mathrm{ae}}(f)$, the Fourier series model provides the best fitting 
	performance and is adopted to characterize the acousto-electric conversion 
	efficiency. Accordingly, the fitted acousto-electric conversion efficiency is 
	modeled as
	\begin{equation}
		\eta_{\mathrm{ae,fit}}(f)
		=
		a_{\mathrm{ae},0}
		+
		\sum_{i=1}^{8}
		\left[
		a_{\mathrm{ae},i}\cos\left(i\omega_{\mathrm{ae}} f\right)
		+
		b_{\mathrm{ae},i}\sin\left(i\omega_{\mathrm{ae}} f\right)
		\right],
		\label{eq:eta_ae_fitting}
	\end{equation}
	where $\omega_{\mathrm{ae}}$ denotes the fundamental angular frequency of the 
	Fourier series model. The parameters $a_{\mathrm{ae},0}$, $a_{\mathrm{ae},i}$, 
	and $b_{\mathrm{ae},i}$ denote the constant term, cosine coefficients, and sine 
	coefficients of the $i$-th harmonic component, respectively. The corresponding 
	fitting curves of $\eta_{\mathrm{ea}}(f)$ and $\eta_{\mathrm{ae}}(f)$ are shown 
	in Fig.~\ref{fig:eta_fitting}.
	
	All fitting parameters are obtained from the calculated efficiency data through 
	the least squares method. The fitted functions $\eta_{\mathrm{ea,fit}}(f)$ and 
	$\eta_{\mathrm{ae,fit}}(f)$ are then used to characterize the frequency-dependent 
	transducer efficiencies at different subcarrier frequencies in the subsequent 
	system model.

\subsection{Signal Model} 
	As stated in Section~\ref{sec:SAIPT MODEL}, the integrated transmitted symbol in 
	the DD domain can be expressed as $x[l,k] = x_\mathrm{I}[l,k] +  x_\mathrm{P}[l,k]$. Next, the ISFFT converts the DD-domain transmitted symbol into the TF domain, which can be expressed as
	\begin{equation}
		\begin{aligned}
			X[m,n]
			&=
			\frac{1}{\sqrt{MN}}
			\sum_{l=0}^{M-1}
			\sum_{k=0}^{N-1}
			x[l,k]
			e^{j2\pi\left(\frac{nk}{N}-\frac{ml}{M}\right)}  \\
			&=
			\frac{1}{\sqrt{MN}}
			\sum_{l=0}^{M-1}
			\sum_{k=0}^{N-1}
			\left(x_{\mathrm{I}}[l,k]+x_{\mathrm{P}}[l,k]\right)
			e^{j2\pi\left(\frac{nk}{N}-\frac{ml}{M}\right)} \\
			&=
			X_{\mathrm{I}}[m,n]+X_{\mathrm{P}}[m,n],
		\end{aligned}
		\label{eq:isfft_transmitted_signal}
	\end{equation}
	where $m=0,1,\ldots,M-1$ and $n=0,1,\ldots,N-1$ denote the subcarrier and 
	time-slot indices, respectively. Here, $X_{\mathrm{I}}[m,n]$ and 
	$X_{\mathrm{P}}[m,n]$ represent the acoustic information transfer (AIT) and acoustic power transfer (APT) components in the TF domain, respectively. The continuous-time baseband waveform $x_{\mathrm{b}}$ is then obtained from $X[m,n]$ through the Heisenberg transform. After frequency up-conversion, $x_{\mathrm{b}}$ is converted into the passband electrical signal $x$, which is further transformed into the transmitted acoustic signal $s$ by the electro-acoustic transducer.
	
	For both the energy receiver and the information receiver, the received TF-domain signal can be expressed in a unified form as
	\begin{equation}
		\begin{aligned}
			Y_q[m,n]
			&=
			H_q[m,n]\sqrt{\eta_{\mathrm{ea},m}}\sqrt{\eta_{\mathrm{ae},m}}X[m,n]
			+N_0[m,n] \\
			&=
			H_q[m,n]\sqrt{\eta_{\mathrm{ea},m}}\sqrt{\eta_{\mathrm{ae},m}}
			\left(X_{\mathrm{I}}[m,n]+X_{\mathrm{P}}[m,n]\right)
			+N_0[m,n],
		\end{aligned}
		\label{eq:tf_received_unified}
	\end{equation}
	where $q\in{\mathrm{P},\mathrm{I}}$ denotes the receiver index, with $\mathrm{P}$ and $\mathrm{I}$ corresponding to the energy receiver and the information receiver, respectively. Here, $H_q[m,n]$ denotes the TF-domain channel coefficient from the transmitter to receiver $q$, $\eta_{\mathrm{ea},m}$ and $\eta_{\mathrm{ae},m}$ denote the electro-acoustic and acousto-electric conversion efficiencies at the $m$-th subcarrier, respectively, and $N_0[m,n]$ denotes the underwater noise component. For notational simplicity, the equivalent TF-domain channel coefficient for receiver $q$ is defined as
	\begin{equation}
		\hat{H}_q[m,n]
		=
		H_q[m,n]\sqrt{\eta_{\mathrm{ea},m}}\sqrt{\eta_{\mathrm{ae},m}}.
		\label{eq:equivalent_channel_unified}
	\end{equation}
	Then, \eqref{eq:tf_received_unified} can be rewritten as
	\begin{equation}
		Y_q[m,n]
		=
		\hat{H}_q[m,n]
		\left(X_{\mathrm{I}}[m,n]+X_{\mathrm{P}}[m,n]\right)
		+N_0[m,n].
		\label{eq:tf_received_equivalent_unified}
	\end{equation}
	
	For the energy receiver, $Y_{\mathrm{P}}[m,n]$ characterizes the received TF-domain signal, whose corresponding time-domain electrical waveform is fed into the rectifier for energy harvesting. For the information receiver, the received acoustic signal $r_{\mathrm{I}}$ is first converted into the electrical signal $y_{\mathrm{I}}$ by the acoustic-electric transducer, and the corresponding TF-domain signal $Y_{\mathrm{I}}[m,n]$ is obtained after frequency down-conversion and the Wigner transform.
	
	Since the power-transfer waveform is deterministic and known at the information 
	receiver, its contribution can be cancelled before information demodulation. 
	Considering imperfect cancellation, the residual power-transfer component is 
	characterized by the residual interference factor $\lambda\in[0,1]$. The 
	remaining TF-domain information signal is then given by
	\begin{equation}
		\widetilde{Y}_{\mathrm{I}}[m,n]
		=
		\hat{H}_{\mathrm{I}}[m,n]
		\left(
		X_{\mathrm{I}}[m,n]
		+
		\sqrt{\lambda} X_{\mathrm{P}}[m,n]
		\right)
		+
		N_0[m,n], 
		\label{eq:tf_information_after_cancellation}
	\end{equation}
	where $\lambda=0$ indicates perfect cancellation of the power-transfer waveform, 
	while $\lambda=1$ corresponds to the case without cancellation.
	After performing the SFFT, the received information signal in the DD domain can 
	be expressed as
	\begin{equation}
		y_{\mathrm{I}}[l,k]
		=
		\frac{1}{\sqrt{MN}}
		\sum_{m=0}^{M-1}
		\sum_{n=0}^{N-1}
		\widetilde{Y}_{\mathrm{I}}[m,n]
		e^{-j2\pi\left(\frac{nk}{N}-\frac{ml}{M}\right)}.
		\label{eq:sfft_information_receiver}
	\end{equation}
	
\section{JOINT OTFS-SAIPT WAVEFORM DESIGN} \label{sec:WAVEFORM DESIGN}
\subsection{Performance Analysis}

\subsubsection{Acoustic Power Transfer Performance}
	For the energy receiver, the received signal $y_\mathrm{P}$ is rectified to yield a DC output, which is subsequently used to
	power the underwater node. 
	According to the nonlinear diode model of the rectifier, the relationship between the DC output and the input signal $y_{\mathrm{P}}$ can be expressed as \cite{clerckxWaveformDesignWireless2016}
	\begin{equation}
		z_{\mathrm{DC}}
		=
		k_2 R_{\mathrm{trans}} \mathbb{E}\left\{\mathcal{P}\left(y_{\mathrm{P}}^2\right)\right\}
		+
		k_4 R_{\mathrm{trans}}^2 \mathbb{E}\left\{\mathcal{P}\left(y_{\mathrm{P}}^4\right)\right\},
		\label{eq:zdc_nonlinear_model}
	\end{equation}
	where $k_2$ and $k_4$ denote the constants determined by the rectifier and $R_{\mathrm{trans}}$ represents the input impedance of the transducer. The variable $z_{\mathrm{DC}}$ is proportional to the output DC power; therefore, maximizing 
	the DC output power is equivalent to maximizing $z_{\mathrm{DC}}$.
	
	Since the received energy signal contains both AIT and 
	APT components, the electrical signal $y_{\mathrm{P}}$ can be 
	written as
	\begin{equation}
		y_{\mathrm{P}}
		=
		y_{\mathrm{P,I}}+y_{\mathrm{P,P}},
		\label{eq:energy_signal_decomposition}
	\end{equation}
	where $y_{\mathrm{P,I}}$ and $y_{\mathrm{P,P}}$ denote the received 
	AIT and APT waveforms at the energy receiver, 
	respectively. 
	
	Since $y_{\mathrm{P,I}}$ is a zero-mean random signal and $y_{\mathrm{P,P}}$ is deterministic, we have
	$\mathbb{E}\!\left[\mathcal{P}(y_{\mathrm{P,P}}y_{\mathrm{P,I}})\right]= 
	\mathbb{E}\!\left[\mathcal{P}(y_{\mathrm{P,P}}^{3}y_{\mathrm{P,I}})\right]= 
	\mathbb{E}\!\left[\mathcal{P}(y_{\mathrm{P,P}}y_{\mathrm{P,I}}^{3})\right]=0$.
	Then, \eqref{eq:zdc_nonlinear_model} can be reformulated as
	
	\begin{equation}
		\begin{aligned}
			z_{\mathrm{DC}}
			&=
			k_2 R_{\mathrm{trans}}
			\mathcal{P}\left(y_{\mathrm{P,P}}^2\right)
			+
			k_2 R_{\mathrm{trans}}
			\mathbb{E}\left\{\mathcal{P}\left(y_{\mathrm{P,I}}^2\right)\right\} \\
			&\quad
			+
			k_4 R_{\mathrm{trans}}^2
			\mathcal{P}\left(y_{\mathrm{P,P}}^4\right)
			+
			k_4 R_{\mathrm{trans}}^2
			\mathbb{E}\left\{\mathcal{P}\left(y_{\mathrm{P,I}}^4\right)\right\} \\
			&\quad
			+
			6k_4 R_{\mathrm{trans}}^2
			\mathcal{P}\left(y_{\mathrm{P,P}}^2\right)
			\mathbb{E}\left\{\mathcal{P}\left(y_{\mathrm{P,I}}^2\right)\right\}.
		\end{aligned}
		\label{eq:zdc_decomposed}
	\end{equation}
	where we have:
	\begin{equation}
		\left\{
		\begin{aligned}
			&\mathcal{P}\!\left(y_{\mathrm{P,P}}^{2}\right)
			=
			\frac{1}{N}
			\sum_{n=0}^{N-1}
			\sum_{m=0}^{M-1}
			\left|
			\hat{H}_{\mathrm{P}}[m,n]X_{\mathrm{P}}[m,n]
			\right|^{2},\\[1mm]
			&\mathbb{E}\!\left\{
			\mathcal{P}\!\left(y_{\mathrm{P,I}}^{2}\right)
			\right\}
			=
			\frac{1}{N}
			\sum_{n=0}^{N-1}
			\sum_{m=0}^{M-1}
			\left|
			\hat{H}_{\mathrm{P}}[m,n]X_{\mathrm{I}}[m,n]
			\right|^{2},\\[1mm]
			&\mathcal{P}\!\left(y_{\mathrm{P,P}}^{4}\right)
			=
			\frac{3}{2N}
			\sum_{n=0}^{N-1}
			\sum_{\substack{m_1,m_2,m_3,m_4\\m_1+m_2=m_3+m_4}}
			\Big(
			\hat{H}_{\mathrm{P}}[m_1,n]X_{\mathrm{P}}[m_1,n]
			\hat{H}_{\mathrm{P}}[m_2,n]X_{\mathrm{P}}[m_2,n]\\
			&\quad\quad\quad\quad\quad\quad\quad\quad\quad\quad\quad\quad\times
			\hat{H}_{\mathrm{P}}^{*}[m_3,n]X_{\mathrm{P}}^{*}[m_3,n]
			\hat{H}_{\mathrm{P}}^{*}[m_4,n]X_{\mathrm{P}}^{*}[m_4,n]
			\Big),\\[1mm]
			&\mathbb{E}\!\left\{
			\mathcal{P}\!\left(y_{\mathrm{P,I}}^{4}\right)
			\right\}
			=
			\frac{3}{N}
			\sum_{n=0}^{N-1}
			\left(
			\sum_{m=0}^{M-1}
			\left|
			\hat{H}_{\mathrm{P}}[m,n]X_{\mathrm{I}}[m,n]
			\right|^{2}
			\right)^2.
		\end{aligned}
		\right.
		\label{eq:matrix_terms}
	\end{equation}
		
	To facilitate the subsequent optimization, define the TF-domain signal vectors at 
	the $n$-th time slot as
	\begin{align}
		\mathbf{x}_{\mathrm{P},n}
		&=
		\left[
		X_{\mathrm{P}}[0,n],
		X_{\mathrm{P}}[1,n],
		\ldots,
		X_{\mathrm{P}}[M-1,n]
		\right]^T,
		\label{eq:x_p_n_vector}
		\\
		\mathbf{x}_{\mathrm{I},n}
		&=
		\left[
		X_{\mathrm{I}}[0,n],
		X_{\mathrm{I}}[1,n],
		\ldots,
		X_{\mathrm{I}}[M-1,n]
		\right]^T.
		\label{eq:x_i_n_vector}
	\end{align}
	
	For the energy receiver, the equivalent channel vector at the $n$-th time slot is 
	defined as
	\begin{equation}
		\hat{\mathbf{h}}_{\mathrm{P},n}
		=
		\left[
		\hat{H}_{\mathrm{P}}[0,n],
		\hat{H}_{\mathrm{P}}[1,n],
		\ldots,
		\hat{H}_{\mathrm{P}}[M-1,n]
		\right]^T.
		\label{eq:h_p_n_vector}
	\end{equation}
	
	Then, the matrix $\mathbf{Q}_{\mathrm{P},n}$ is defined as
	\begin{equation}
		\mathbf{Q}_{\mathrm{P},n}
		=
		\hat{\mathbf{h}}_{\mathrm{P},n}^*
		\hat{\mathbf{h}}_{\mathrm{P},n}^{T}.
		\label{eq:q_p_n}
	\end{equation}
	
	We further introduce a family of auxiliary matrices
	$\mathbf{Q}_{\mathrm{P},n,c}$. Specifically, $\mathbf{Q}_{\mathrm{P},n,c}$ is an $M\times M$ matrix obtained by preserving only a selected subset of the entries of $\mathbf{Q}_{\mathrm{P},n}$, with all other
	entries set to zero. For $c \in \{1,\ldots,M-1\}$, the matrix $\mathbf{Q}_{\mathrm{P},n,c}$ retains
	the elements located on the $c$-th super-diagonal of $\mathbf{Q}_{\mathrm{P},n}$; and for $c=0$, it preserves the main diagonal of $\mathbf{Q}_{\mathrm{P},n}$.
	
	 Accordingly, the terms in 
	\eqref{eq:zdc_decomposed} can be rewritten as
	\begin{equation}
		\left\{
		\begin{aligned}
			\mathcal{P}\!\left(y_{\mathrm{P,P}}^{2}\right)
			&=
			\frac{1}{N}
			\sum_{n=0}^{N-1}
			\mathbf{x}_{\mathrm{P},n}^{H}
			\mathbf{Q}_{\mathrm{P},n,0}
			\mathbf{x}_{\mathrm{P},n},\\[1mm]
			\mathbb{E}\!\left\{
			\mathcal{P}\!\left(y_{\mathrm{P,I}}^{2}\right)
			\right\}
			&=
			\frac{1}{N}
			\sum_{n=0}^{N-1}
			\mathbf{x}_{\mathrm{I},n}^{H}
			\mathbf{Q}_{\mathrm{P},n,0}
			\mathbf{x}_{\mathrm{I},n},\\[1mm]
			\mathcal{P}\!\left(y_{\mathrm{P,P}}^{4}\right)
			&=
			\frac{3}{2N}
			\sum_{n=0}^{N-1}
			\left[
			\left(
			\mathbf{x}_{\mathrm{P},n}^{H}
			\mathbf{Q}_{\mathrm{P},n,0}
			\mathbf{x}_{\mathrm{P},n}
			\right)^2
			+
			2\sum_{c=1}^{M-1}
			\left|
			\mathbf{x}_{\mathrm{P},n}^{H}
			\mathbf{Q}_{\mathrm{P},n,c}
			\mathbf{x}_{\mathrm{P},n}
			\right|^2
			\right],\\[1mm]
			\mathbb{E}\!\left\{
			\mathcal{P}\!\left(y_{\mathrm{P,I}}^{4}\right)
			\right\}
			&=
			\frac{3}{N}
			\sum_{n=0}^{N-1}
			\left(
			\mathbf{x}_{\mathrm{I},n}^{H}
			\mathbf{Q}_{\mathrm{P},n,0}
			\mathbf{x}_{\mathrm{I},n}
			\right)^2.
		\end{aligned}
		\right.
		\label{eq:matrix_power_terms}
	\end{equation}
	
	Substituting \eqref{eq:matrix_power_terms} into 
	\eqref{eq:zdc_decomposed}, the DC output at the energy receiver can be expressed 
	as
	\begin{equation}
		\begin{aligned}
			&z_{\mathrm{DC}}
			\left(
			\left\{\mathbf{x}_{\mathrm{P},n}\right\}_{n=0}^{N-1},
			\left\{\mathbf{x}_{\mathrm{I},n}\right\}_{n=0}^{N-1}
			\right)
			\\&=
			k_2 R_{\mathrm{trans}}
			\Bigg[
			\frac{1}{N}
			\sum_{n=0}^{N-1}
			\mathbf{x}_{\mathrm{P},n}^{H}
			\mathbf{Q}_{\mathrm{P},n,0}
			\mathbf{x}_{\mathrm{P},n}
			+
			\frac{1}{N}
			\sum_{n=0}^{N-1}
			\mathbf{x}_{\mathrm{I},n}^{H}
			\mathbf{Q}_{\mathrm{P},n,0}
			\mathbf{x}_{\mathrm{I},n}
			\Bigg]
			\\
			&
			+
			k_4 R_{\mathrm{trans}}^2
			\Bigg[
			\frac{3}{2N}
			\sum_{n=0}^{N-1}
			\left(
			\left(
			\mathbf{x}_{\mathrm{P},n}^{H}
			\mathbf{Q}_{\mathrm{P},n,0}
			\mathbf{x}_{\mathrm{P},n}
			\right)^2
			\right.
			\left.
			+
			2\sum_{c=1}^{M-1}
			\left|
			\mathbf{x}_{\mathrm{P},n}^{H}
			\mathbf{Q}_{\mathrm{P},n,c}
			\mathbf{x}_{\mathrm{P},n}
			\right|^2
			\right)
			\\
			&\qquad
			+
			\frac{6}{N^2}
			\left(
			\sum_{n=0}^{N-1}
			\mathbf{x}_{\mathrm{P},n}^{H}
			\mathbf{Q}_{\mathrm{P},n,0}
			\mathbf{x}_{\mathrm{P},n}
			\right)
			\left(
			\sum_{n=0}^{N-1}
			\mathbf{x}_{\mathrm{I},n}^{H}
			\mathbf{Q}_{\mathrm{P},n,0}
			\mathbf{x}_{\mathrm{I},n}
			\right)
			\\
			&\qquad
			+
			\frac{3}{N}
			\sum_{n=0}^{N-1}
			\left(
			\mathbf{x}_{\mathrm{I},n}^{H}
			\mathbf{Q}_{\mathrm{P},n,0}
			\mathbf{x}_{\mathrm{I},n}
			\right)^2
			\Bigg].
		\end{aligned}
		\label{eq:zdc_complex_matrix_form}
	\end{equation}
	
	Furthermore, $\mathbf{x}_{\mathrm{P},n}$, $\mathbf{x}_{\mathrm{I},n}$ and 
	$\mathbf{Q}_{\mathrm{P},n,c}$ can be decomposed into their real and imaginary parts, 
	thereby converting the expressions from the complex domain to the real domain and 
	reducing the complexity of subsequent optimization. Accordingly, the DC output can be equivalently rewritten as
	\begin{equation}
		\begin{aligned}
			&z_{\mathrm{DC}}
			\left(
			\left\{\hat{\mathbf{x}}_{\mathrm{P},n}\right\}_{n=0}^{N-1},
			\left\{\hat{\mathbf{x}}_{\mathrm{I},n}\right\}_{n=0}^{N-1}
			\right)\\
			&=
			k_2 R_{\mathrm{trans}}
			\Bigg[
			\frac{1}{N}
			\sum_{n=0}^{N-1}
			\hat{\mathbf{x}}_{\mathrm{P},n}^{T}
			\mathbf{T}_{\mathrm{P},n,0}
			\hat{\mathbf{x}}_{\mathrm{P},n}
			+
			\frac{1}{N}
			\sum_{n=0}^{N-1}
			\hat{\mathbf{x}}_{\mathrm{I},n}^{T}
			\mathbf{T}_{\mathrm{P},n,0}
			\hat{\mathbf{x}}_{\mathrm{I},n}
			\Bigg]
			\\
			&\quad
			+
			k_4 R_{\mathrm{trans}}^2
			\Bigg[
			\frac{3}{2N}
			\sum_{n=0}^{N-1}
			\left(
			\left(
			\hat{\mathbf{x}}_{\mathrm{P},n}^{T}
			\mathbf{T}_{\mathrm{P},n,0}
			\hat{\mathbf{x}}_{\mathrm{P},n}
			\right)^2
			\right.
			\left.
			+
			2\sum_{c=1}^{M-1}
			\left(
			\hat{\mathbf{x}}_{\mathrm{P},n}^{T}
			\mathbf{T}_{\mathrm{P},n,c}
			\hat{\mathbf{x}}_{\mathrm{P},n}
			\right)^2
			\right)
			\\
			&\qquad
			+
			\frac{6}{N^2}
			\left(
			\sum_{n=0}^{N-1}
			\hat{\mathbf{x}}_{\mathrm{P},n}^{T}
			\mathbf{T}_{\mathrm{P},n,0}
			\hat{\mathbf{x}}_{\mathrm{P},n}
			\right)
			\left(
			\sum_{n=0}^{N-1}
			\hat{\mathbf{x}}_{\mathrm{I},n}^{T}
			\mathbf{T}_{\mathrm{P},n,0}
			\hat{\mathbf{x}}_{\mathrm{I},n}
			\right)
			\\
			&\qquad
			+
			\frac{3}{N}
			\sum_{n=0}^{N-1}
			\left(
			\hat{\mathbf{x}}_{\mathrm{I},n}^{T}
			\mathbf{T}_{\mathrm{P},n,0}
			\hat{\mathbf{x}}_{\mathrm{I},n}
			\right)^2
			\Bigg],
		\end{aligned}
		\label{eq:zdc_real_matrix_form}
	\end{equation}
	where
	$\hat{\mathbf{x}}_{\mathrm{P},n}
	\triangleq
	\begin{bmatrix}
		\Re\{\mathbf{x}_{\mathrm{P},n}\} \\
		\Im\{\mathbf{x}_{\mathrm{P},n}\}
	\end{bmatrix}
	\in\mathbb{R}^{2M}$,
	$\hat{\mathbf{x}}_{\mathrm{I},n}
	\triangleq
	\begin{bmatrix}
		\Re\{\mathbf{x}_{\mathrm{I},n}\} \\
		\Im\{\mathbf{x}_{\mathrm{I},n}\}
	\end{bmatrix}
	\in\mathbb{R}^{2M}$,
	and
	$
		\mathbf{T}_{\mathrm{P},n,c}
		\triangleq
		\begin{bmatrix}
			\Re\{\mathbf{Q}_{\mathrm{P},n,c}\} & -\Im\{\mathbf{Q}_{\mathrm{P},n,c}\} \\
			\Im\{\mathbf{Q}_{\mathrm{P},n,c}\} & \Re\{\mathbf{Q}_{\mathrm{P},n,c}\}
		\end{bmatrix}
		\in \mathbb{R}^{2M\times 2M}.
		\label{eq:r_p_n_c_definition}
	$

\subsubsection{Acoustic Information Transfer Performance}
	For the information receiver, the achievable data rate can be expressed according 
	to Shannon's capacity formula as
	\begin{equation}
			R
			=
			\sum_{m=0}^{M-1}
			\Delta f
			\log_{2}
			\left(
			1+
			\frac{
				\sum_{n=0}^{N-1}
				\left|
				\hat{H}_{\mathrm{I}}[m,n]X_{\mathrm{I}}[m,n]
				\right|^{2}
			}{
				\lambda
				\sum_{n=0}^{N-1}
				\left|
				\hat{H}_{\mathrm{I}}[m,n]X_{\mathrm{P}}[m,n]
				\right|^{2}
				+
				P_{\mathrm{noise},m}
				+
				P_{\mathrm{cov}}
			}
			\right),
		\label{eq:achievable_rate}
	\end{equation}
	where $\Delta f$ denotes the subcarrier spacing, $P_{\mathrm{noise},m}$ denotes 
	the underwater acoustic noise power at the $m$-th subcarrier, and $P_{\mathrm{cov}}$ 
	denotes the noise power introduced by the down-conversion process.
\subsection{Problem Formulation}
	The objective of this work is to jointly design the DD-domain information-transfer 
	symbols $x_{\mathrm{I}}[l,k]$ and power-transfer symbols $x_{\mathrm{P}}[l,k]$ 
	under a given transmit power budget $P_{\mathrm{tx}}$. Specifically, the DC output 
	$z_{\mathrm{DC}}$ at the energy-harvesting circuit is maximized while satisfying 
	the minimum achievable data rate requirement $R_{\min}$ of the information receiver.
	
	Since both the achievable rate $R$ and the DC output $z_{\mathrm{DC}}$ have been 
	formulated in the TF domain, the optimization variables are equivalently transformed 
	from the DD-domain symbols $x_{\mathrm{I}}[l,k]$ and $x_{\mathrm{P}}[l,k]$ to the 
	TF-domain symbols $X_{\mathrm{I}}[m,n]$ and $X_{\mathrm{P}}[m,n]$. Accordingly, the 
	optimization is performed over the real-valued vector sets 
	$\{\hat{\mathbf{x}}_{\mathrm{P},n}\}_{n=0}^{N-1}$ and 
	$\{\hat{\mathbf{x}}_{\mathrm{I},n}\}_{n=0}^{N-1}$. After obtaining the optimized 
	TF-domain variables, the corresponding DD-domain symbols can be recovered through 
	the SFFT operation.
	
	Therefore, the optimization problem can be formulated as
	\begin{subequations}
		\label{eq:problem_p1}
		\begin{align}
			(\mathrm{P}1):\quad
			\max_{\left\{\hat{\mathbf{x}}_{\mathrm{P},n}\right\},
				\left\{\hat{\mathbf{x}}_{\mathrm{I},n}\right\}}
			\quad
			& z_{\mathrm{DC}}(\left\{\hat{\mathbf{x}}_{\mathrm{P},n}\right\},
			\left\{\hat{\mathbf{x}}_{\mathrm{I},n}\right\})
			\label{eq:p1_objective}
			\\
			\mathrm{s.t.}\quad
			& R \geq R_{\min},
			\label{eq:p1_rate_constraint}
			\\
			&
			\frac{1}{N}
			\sum_{n=0}^{N-1}
			\left(
			\left\|
			\hat{\mathbf{x}}_{\mathrm{P},n}
			\right\|_{2}^{2}
			+
			\left\|
			\hat{\mathbf{x}}_{\mathrm{I},n}
			\right\|_{2}^{2}
			\right)
			\leq
			P_{\mathrm{tx}}.
			\label{eq:p1_power_constraint}
		\end{align}
	\end{subequations}
\subsection{Joint Waveform Design}
	Since problem (P1) is non-convex due to the nonlinear objective function and the
	achievable-rate constraint, the successive convex approximation (SCA) method is
	adopted to obtain a tractable solution. Specifically, at each iteration, the
	non-convex terms are approximated by their first-order Taylor expansions around
	the feasible point obtained from the previous iteration.
	
	Let
	$\{\hat{\mathbf{x}}_{\mathrm{P},n}^{(r)}\}$ and
	$\{\hat{\mathbf{x}}_{\mathrm{I},n}^{(r)}\}$ denote the feasible point obtained at
	the $r$-th iteration. Then, the objective function $z_{\mathrm{DC}}$ can be
	approximated as
	
	\begin{equation}
		\begin{aligned}
			\widetilde{z}_{\mathrm{DC}}
			&\left(
			\left\{\hat{\mathbf{x}}_{\mathrm{P},n}\right\},
			\left\{\hat{\mathbf{x}}_{\mathrm{I},n}\right\};
			\left\{\hat{\mathbf{x}}_{\mathrm{P},n}^{(r)}\right\},
			\left\{\hat{\mathbf{x}}_{\mathrm{I},n}^{(r)}\right\}
			\right)
			\\
			&\approx
			z_{\mathrm{DC}}
			\left(
			\left\{\hat{\mathbf{x}}_{\mathrm{P},n}^{(r)}\right\},
			\left\{\hat{\mathbf{x}}_{\mathrm{I},n}^{(r)}\right\}
			\right)
			\\
			&\quad
			+
			\sum_{n=0}^{N-1}
			\nabla_{\hat{\mathbf{x}}_{\mathrm{P},n}}
			z_{\mathrm{DC}}
			\left(
			\left\{\hat{\mathbf{x}}_{\mathrm{P},n}^{(r)}\right\},
			\left\{\hat{\mathbf{x}}_{\mathrm{I},n}^{(r)}\right\}
			\right)^{T}
			\left(
			\hat{\mathbf{x}}_{\mathrm{P},n}
			-
			\hat{\mathbf{x}}_{\mathrm{P},n}^{(r)}
			\right)
			\\
			&\quad
			+
			\sum_{n=0}^{N-1}
			\nabla_{\hat{\mathbf{x}}_{\mathrm{I},n}}
			z_{\mathrm{DC}}
			\left(
			\left\{\hat{\mathbf{x}}_{\mathrm{P},n}^{(r)}\right\},
			\left\{\hat{\mathbf{x}}_{\mathrm{I},n}^{(r)}\right\}
			\right)^{T}
			\left(
			\hat{\mathbf{x}}_{\mathrm{I},n}
			-
			\hat{\mathbf{x}}_{\mathrm{I},n}^{(r)}
			\right).
		\end{aligned}
		\label{eq:zdc_taylor_approximation}
	\end{equation}
	where
	$\nabla_{\hat{\mathbf{x}}_{\mathrm{P},n}}z_{\mathrm{DC}}(\cdot)$ and
	$\nabla_{\hat{\mathbf{x}}_{\mathrm{I},n}}z_{\mathrm{DC}}(\cdot)$ denote the gradients
	of $z_{\mathrm{DC}}$ with respect to $\hat{\mathbf{x}}_{\mathrm{P},n}$ and
	$\hat{\mathbf{x}}_{\mathrm{I},n}$, respectively, evaluated at the feasible point of the $r$-th iteration.
	
	Consequently, problem (P1) can be transformed into
	\begin{subequations}
		\label{eq:problem_p2}
		\begin{align}
			(\mathrm{P}2):\quad
			\max_{\left\{\hat{\mathbf{x}}_{\mathrm{P},n}\right\},
				\left\{\hat{\mathbf{x}}_{\mathrm{I},n}\right\}}
			\quad
			&
			\widetilde{z}_{\mathrm{DC}}
			\left(
			\left\{\hat{\mathbf{x}}_{\mathrm{P},n}\right\},
			\left\{\hat{\mathbf{x}}_{\mathrm{I},n}\right\};
			\left\{\hat{\mathbf{x}}_{\mathrm{P},n}^{(r)}\right\},
			\left\{\hat{\mathbf{x}}_{\mathrm{I},n}^{(r)}\right\}
			\right)
			\label{eq:p2_objective}
			\\
			\mathrm{s.t.}\quad
			&
			R \geq R_{\min},
			\label{eq:p2_rate_constraint}
			\\
			&
			\frac{1}{N}
			\sum_{n=0}^{N-1}
			\left(
			\left\|
			\hat{\mathbf{x}}_{\mathrm{P},n}
			\right\|_{2}^{2}
			+
			\left\|
			\hat{\mathbf{x}}_{\mathrm{I},n}
			\right\|_{2}^{2}
			\right)
			\leq
			P_{\mathrm{tx}}.
			\label{eq:p2_power_constraint}
		\end{align}
	\end{subequations}
	
	To handle the non-convex achievable-rate constraint \eqref{eq:p2_rate_constraint}, 
	the following two auxiliary functions are defined for each subcarrier index $m$ as
	\begin{equation}
		\begin{aligned}
			U_m
			&=
			\sum_{n=0}^{N-1}
			\left|
			\hat{H}_{\mathrm{I}}[m,n]X_{\mathrm{I}}[m,n]
			\right|^2
			+
			\lambda
			\sum_{n=0}^{N-1}
			\left|
			\hat{H}_{\mathrm{I}}[m,n]X_{\mathrm{P}}[m,n]
			\right|^2\\
			&~~~+
			P_{\mathrm{noise},m}
			+
			P_{\mathrm{cov}},
			\\
			V_m
			&=
			\lambda
			\sum_{n=0}^{N-1}
			\left|
			\hat{H}_{\mathrm{I}}[m,n]X_{\mathrm{P}}[m,n]
			\right|^2
			+
			P_{\mathrm{noise},m}
			+
			P_{\mathrm{cov}}.
		\end{aligned}
		\label{eq:U_V_definition}
	\end{equation}
	
	Then, auxiliary variables $\{s_m\}$ and $\{t_m\}$ are introduced as
	\begin{equation}
		e^{s_m}=U_m,\quad e^{t_m}=V_m,
		\quad m=0,1,\ldots,M-1.
		\label{eq:auxiliary_variables_rate}
	\end{equation}

	Therefore, problem (P2) can be further reformulated as
	\begin{subequations}
		\label{eq:problem_p3}
		\begin{align}
			(\mathrm{P}3):\quad 
			&\max_{\left\{\hat{\mathbf{x}}_{\mathrm{P},n}\right\},
				\left\{\hat{\mathbf{x}}_{\mathrm{I},n}\right\},
				\{s_m\},\{t_m\}} 
			\quad \widetilde{z}_{\mathrm{DC}}
			\left(
			\left\{\hat{\mathbf{x}}_{\mathrm{P},n}\right\},
			\left\{\hat{\mathbf{x}}_{\mathrm{I},n}\right\};
			\left\{\hat{\mathbf{x}}_{\mathrm{P},n}^{(r)}\right\},
			\left\{\hat{\mathbf{x}}_{\mathrm{I},n}^{(r)}\right\}
			\right)
			\label{eq:p3_objective}
			\\
			\mathrm{s.t.}\quad
			&
			\Delta f
			\sum_{m=0}^{M-1}
			\left(s_m-t_m\right)\log_2 e
			\geq R_{\min},
			\label{eq:p3_rate_constraint}
			\\
			&
			U_m \geq e^{s_m},
			\quad \forall m,
			\label{eq:p3_s_constraint}
			\\
			&
			V_m \leq e^{t_m},
			\quad \forall m,
			\label{eq:p3_t_constraint}
			\\
			&
			\frac{1}{N}
			\sum_{n=0}^{N-1}
			\left(
			\left\|
			\hat{\mathbf{x}}_{\mathrm{P},n}
			\right\|_2^2
			+
			\left\|
			\hat{\mathbf{x}}_{\mathrm{I},n}
			\right\|_2^2
			\right)
			\leq P_{\mathrm{tx}}.
			\label{eq:p3_power_constraint}
		\end{align}
	\end{subequations}
	
	By further performing a first-order Taylor expansion of
	\eqref{eq:p3_s_constraint} and \eqref{eq:p3_t_constraint}, problem (P3) can be reformulated as
	\begin{subequations}
		\label{eq:problem_p4}
		\begin{align}
			(\mathrm{P}4):\quad 
			&\max_{\left\{\hat{\mathbf{x}}_{\mathrm{P},n}\right\},
				\left\{\hat{\mathbf{x}}_{\mathrm{I},n}\right\},
				\{s_m\},\{t_m\}}			
			\quad \widetilde{z}_{\mathrm{DC}}
			\left(
			\left\{\hat{\mathbf{x}}_{\mathrm{P},n}\right\},
			\left\{\hat{\mathbf{x}}_{\mathrm{I},n}\right\};
			\left\{\hat{\mathbf{x}}_{\mathrm{P},n}^{(r)}\right\},
			\left\{\hat{\mathbf{x}}_{\mathrm{I},n}^{(r)}\right\}
			\right)
			\label{eq:p4_objective}
			\\
			\mathrm{s.t.}\quad
			&
			\Delta f
			\sum_{m=0}^{M-1}
			\left(s_m-t_m\right)\log_2 e
			\geq R_{\min},
			\label{eq:p4_rate_constraint}
			\\
			&
			\widetilde{U}_m^{(r)}
			\geq e^{s_m},
			\quad \forall m,
			\label{eq:p4_s_constraint}
			\\
			&
			V_m
			\leq
			e^{t_m^{(r)}}
			+
			e^{t_m^{(r)}}
			\left(
			t_m-t_m^{(r)} \right),
			\quad \forall m,
			\label{eq:p4_t_constraint}
			\\
			&
			\frac{1}{N}
			\sum_{n=0}^{N-1}
			\left(
			\left\|
			\hat{\mathbf{x}}_{\mathrm{P},n}
			\right\|_2^2
			+
			\left\|
			\hat{\mathbf{x}}_{\mathrm{I},n}
			\right\|_2^2
			\right)
			\leq P_{\mathrm{tx}}.
			\label{eq:p4_power_constraint}
		\end{align}
	\end{subequations}
	where $\left\{t_m^{(r)}\right\}$ denotes  the value of $t_m$ at the $r$-th iteration. Moreover, $\widetilde{U}_m^{(r)}$ is the first-order Taylor 
	approximation of $U_m$ at 
	$\left\{\hat{\mathbf{x}}_{\mathrm{P},n}^{(r)}\right\}$ and 
	$\left\{\hat{\mathbf{x}}_{\mathrm{I},n}^{(r)}\right\}$, which is given by
	\begin{equation}
		\begin{aligned}
			\widetilde{U}_m^{(r)}
			&=
			U_m^{(r)}
			+
			\sum_{n=0}^{N-1}
			\nabla_{\hat{\mathbf{x}}_{\mathrm{P},n}}
			U_m^{(r)T}
			\left(
			\hat{\mathbf{x}}_{\mathrm{P},n}
			-
			\hat{\mathbf{x}}_{\mathrm{P},n}^{(r)}
			\right)
			\\
			&\quad
			+
			\sum_{n=0}^{N-1}
			\nabla_{\hat{\mathbf{x}}_{\mathrm{I},n}}
			U_m^{(r)T}
			\left(
			\hat{\mathbf{x}}_{\mathrm{I},n}
			-
			\hat{\mathbf{x}}_{\mathrm{I},n}^{(r)}
			\right),
		\end{aligned}
		\label{eq:U_taylor}
	\end{equation}
	where $U_m^{(r)}$ denotes the value of $U_m$ at the $r$-th iteration, and
	$\nabla_{\hat{\mathbf{x}}_{\mathrm{P},n}} U_m^{(r)}$ and
	$\nabla_{\hat{\mathbf{x}}_{\mathrm{I},n}} U_m^{(r)}$ denote the gradients of
	$U_m$ with respect to $\hat{\mathbf{x}}_{\mathrm{P},n}$ and
	$\hat{\mathbf{x}}_{\mathrm{I},n}$ evaluated at
	$\hat{\mathbf{x}}_{\mathrm{P},n}^{(r)}$ and
	$\hat{\mathbf{x}}_{\mathrm{I},n}^{(r)}$, respectively.
	
	Consequently, by applying the rate transformation to the achievable-rate constraint 
	and performing first-order Taylor expansions of both the objective function and the 
	non-convex constraints, the TF-domain waveform vectors can be optimized via the 
	successive convex approximation (SCA) method by iteratively solving problem (P4). 
	At the $r$-th iteration, the first-order Taylor expansion points 
	$\left\{\hat{\mathbf{x}}_{\mathrm{P},n}^{(r)}\right\}$ and 
	$\left\{\hat{\mathbf{x}}_{\mathrm{I},n}^{(r)}\right\}$ are updated according to the 
	waveform vectors obtained from the previous iteration. The initial expansion points 
	are randomly generated feasible vectors satisfying the constraints.
	
	In addition, to enhance the global search capability and reduce the possibility 
	of convergence to a poor local optimum, a multiple-initial-points strategy is 
	adopted. Specifically, several randomly generated feasible symbol vectors are 
	used as the initial expansion points of the SCA procedure. For each initialization, 
	problem (P4) is iteratively solved, and the solution that yields the maximum 
	$z_{\mathrm{DC}}$ is selected as the final result. The detailed procedure is 
	summarized in Algorithm~\ref{alg:OTFS_opt}.

	\begin{algorithm}[hb!]
	\caption{Algorithm for Optimization in OTFS-SAIPT System}
	\label{alg:OTFS_opt}
	\begin{algorithmic}[1]
		\REQUIRE System parameters and constraints; required accuracy $\epsilon$; number of initial points $M_{\mathrm{init}}$
		\ENSURE Optimal waveform vectors $\{\hat{\mathbf{x}}_{\mathrm{P},n}^{\mathrm{opt}}\}$ and $\{\hat{\mathbf{x}}_{\mathrm{I},n}^{\mathrm{opt}}\}$; optimal DC output $z_{\mathrm{DC}}^{\mathrm{opt}}$
		\STATE $z_{\mathrm{DC}}^{\mathrm{opt}} \gets -\infty$
		\FOR{$k = 1$ \TO $M_{\mathrm{init}}$}
		\STATE Randomly generate feasible initial waveform vectors $\{\hat{\mathbf{x}}_{\mathrm{P},n}^{(0)}\}$ and $\{\hat{\mathbf{x}}_{\mathrm{I},n}^{(0)}\}$
		\STATE $z_{\mathrm{DC}}^{(0)} \gets z_{\mathrm{DC}}\left(\{\hat{\mathbf{x}}_{\mathrm{P},n}^{(0)}\},\{\hat{\mathbf{x}}_{\mathrm{I},n}^{(0)}\}\right)$
		\STATE $r \gets 0$
		\REPEAT
		\STATE $r \gets r + 1$
		\STATE Set the Taylor expansion points as $\{\hat{\mathbf{x}}_{\mathrm{P},n}^{(r-1)}\}$ and $\{\hat{\mathbf{x}}_{\mathrm{I},n}^{(r-1)}\}$
		\STATE Obtain $\{\hat{\mathbf{x}}_{\mathrm{P},n}^{(r)}\}$ and $\{\hat{\mathbf{x}}_{\mathrm{I},n}^{(r)}\}$ by solving problem (P4)
		\STATE $z_{\mathrm{DC}}^{(r)} \gets z_{\mathrm{DC}}\left(\{\hat{\mathbf{x}}_{\mathrm{P},n}^{(r)}\},\{\hat{\mathbf{x}}_{\mathrm{I},n}^{(r)}\}\right)$
		\UNTIL $\frac{\left|z_{\mathrm{DC}}^{(r)}-z_{\mathrm{DC}}^{(r-1)}\right|}{\left|z_{\mathrm{DC}}^{(r-1)}\right|}<\epsilon$
		\IF{$z_{\mathrm{DC}}^{(r)} > z_{\mathrm{DC}}^{\mathrm{opt}}$}
		\STATE $z_{\mathrm{DC}}^{\mathrm{opt}} \gets z_{\mathrm{DC}}^{(r)}$
		\STATE $\{\hat{\mathbf{x}}_{\mathrm{P},n}^{\mathrm{opt}}\} \gets \{\hat{\mathbf{x}}_{\mathrm{P},n}^{(r)}\}$
		\STATE $\{\hat{\mathbf{x}}_{\mathrm{I},n}^{\mathrm{opt}}\} \gets \{\hat{\mathbf{x}}_{\mathrm{I},n}^{(r)}\}$
		\ENDIF
		\ENDFOR
	\end{algorithmic}
	\end{algorithm}

\subsection{Convergence Analysis}
	The convergence is guaranteed by the monotonic improvement of the DC output and the boundedness of the feasible region. At each iteration, the 
	original non-convex problem is approximated by a convex subproblem based on the 
	current feasible point. The constructed surrogate objective is tight at the current point, which guarantees the consistency between the surrogate problem and the original problem at the expansion point. Meanwhile, the rate constraint is replaced by a conservative convex approximation, so that the solution obtained at each iteration remains feasible for the original problem.
	
	Since the current feasible point is also feasible for the convex subproblem in the 
	next iteration, solving the subproblem yields an objective value no smaller than that 
	obtained in the previous iteration. Hence, the DC output generated by the proposed 
	algorithm is monotonically non-decreasing. Moreover, due to the transmit power 
	constraint, the DC output is upper bounded. Consequently, the objective sequence 
	generated by the proposed algorithm is guaranteed to converge.
	
	By updating the expansion point with the solution obtained from the previous iteration, 
	the waveform vectors are iteratively improved until the stopping criterion is satisfied. 
	Following the standard convergence results of the SCA framework in \cite{razaviyayn2014successive}, any limit point 
	of the generated sequence is a stationary point of problem (P4). 
	
\subsection{Computational Complexity Analysis}
	We now evaluate the computational complexity of the proposed algorithm. 
	Let $M_{\mathrm{init}}$ denote the number of initial points and $I_{\mathrm{SCA}}$ 
	denote the number of iterations required by the SCA procedure. Problem (P4), which can be formulated as a second-order cone programming (SOCP) problem, is solved via the interior-point method implemented in CVX. According to~\cite{i.p’olikInteriorPointMethods2010}, 
	the complexity of solving SOCP can be expressed as
	$
	\mathcal{O}\!\left(\sqrt{k}\,\log \frac{1}{\tau}
	\left(m^{3}+ m^{2}n + \sum_{i=1}^{k}n_{i}^{2}\right)\right)
	$ where $k$ is the number of cones, $\tau$ is the solution accuracy, 
	$m$ is the number of equality constraints, $n$ is the total dimension 
	of the second-order cone variables, and $n_i$ denotes the dimension 
	of the $i$-th cone. (Note: all quantities above are derived under the SOCP 
	standard form.) Accordingly, the approximate complexity of solving Problem (P4) can be written as
	$
	\mathcal{O}\!\left( \log\frac{1}{\tau}(MN)^{3.5}\right)
	$. Therefore, the overall computational complexity of 
	Algorithm~\ref{alg:OTFS_opt} is given by
	$
	\mathcal{O}\!\left( M_{\mathrm{init}}I_{\mathrm{SCA}}(MN)^{3.5}\log\frac{1}{\tau}\right)
	$, where $M$ and $N$ denote the numbers of subcarriers and time slots, respectively.
	
\section{NUMERICAL RESULTS} \label{NUMERICAL RESULTS}
	\begin{figure}[!t]
	\centering
	\includegraphics[width=0.95\columnwidth]{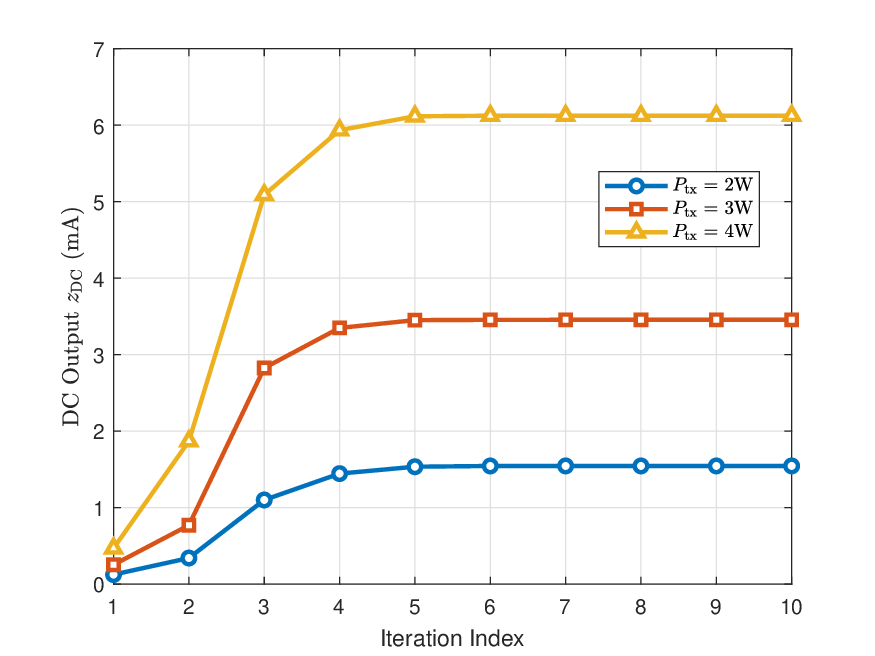}
	\caption{Convergence performance of the proposed SCA algorithm with $M=10$, $N=10$, $d_{\mathrm{I}}=3~\mathrm{m}$, $d_{\mathrm{P}}=4~\mathrm{m}$, $\Delta f=200~\mathrm{Hz}$, $R_{\min}=1000~\mathrm{bps}$, and $\lambda=0.02$.}
	\label{fig:Convergence_curve}
	\end{figure}
	In this section, simulation results are presented to evaluate the DC output performance of the proposed OTFS-based SAIPT waveform design. For the transducer and rectifier, the parameters are set as $k_2=0.0034$, $k_4=0.3859$, and $R_{\mathrm{trans}}=50~\Omega$. Unless otherwise specified, the number of 
	propagation paths is set to $N_{\mathrm{path}}=5$, while $N_{\mathrm{path}}=3$ is also considered in some 
	simulation cases to evaluate the performance under different multipath conditions. Table \ref{tab:parameter_settings} lists the other simulation parameters. 
	
	\begin{table}[h]
		\centering
		\caption{Simulation parameters}
		\label{tab:parameter_settings}
		\begin{tabular}{lc}
			\toprule
			\textbf{Parameters} & \textbf{Values} \\
			\midrule
			Center frequency of transducer, $f_r$ & $30.9~\mathrm{kHz}$ \\
			Underwater sound speed, $c$ & $1500~\mathrm{m/s}$\\
			Maximum velocity of AUV, $v_\mathrm{max}$ & $3~\mathrm{m/s}$\\
			Overall efficiency of the electrical circuitry, $\Phi$ & 0.8 \\
			Down-conversion induced noise, $\sigma_{\mathrm{cov}}^2$ & $-80~\mathrm{dBm}$ \\
			Number of initial points, $M_{\mathrm{init}}$ & 200\\
			\bottomrule
		\end{tabular}
	\end{table}

	Fig.~\ref{fig:Convergence_curve} illustrates the convergence behavior of the proposed SCA-based algorithm under different transmit power levels. It can be observed that the DC output $z_{\mathrm{DC}}$ increases rapidly during the first several iterations for all considered cases and then gradually converges to a stable value. This indicates that the proposed waveform optimization can effectively improve the energy-harvesting performance within a small number of iterations. In addition, a higher transmit power leads to a larger converged DC output, since more transmit power can be exploited by the energy receiver through nonlinear rectification. The convergence trends under different transmit power levels verify the effectiveness and numerical stability of the proposed SCA-based waveform design.
	
	\begin{figure*}[htbp]
		\begin{minipage}{0.48\textwidth}
			\centering
			\includegraphics[width=\linewidth]{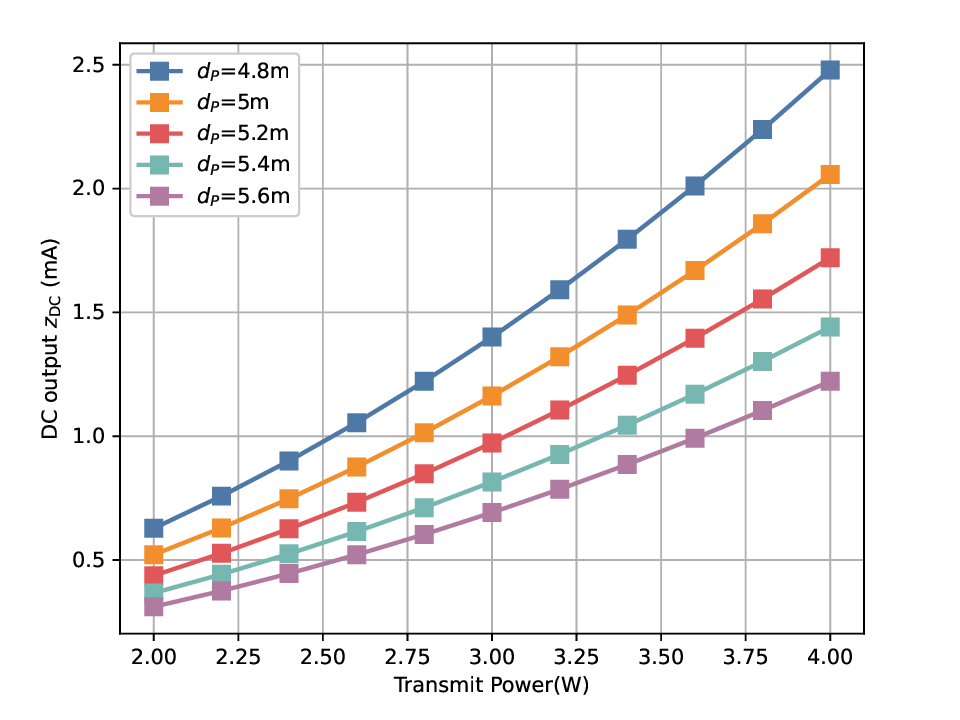}
			\vspace{-25pt}
			\caption{DC output versus transmit power $P_{\text{tx}}$ at $M=10$, $N=10$,
				$d_{\mathrm{I}}=3~\mathrm{m}$, $\Delta f=200~\mathrm{Hz}$, $R_{\min}=1000~\mathrm{bps}$ and $\lambda=0.02$.}
			\label{fig:P_dP}
		\end{minipage}\hfill
		\begin{minipage}{0.48\textwidth}
			\centering
			\includegraphics[width=\linewidth]{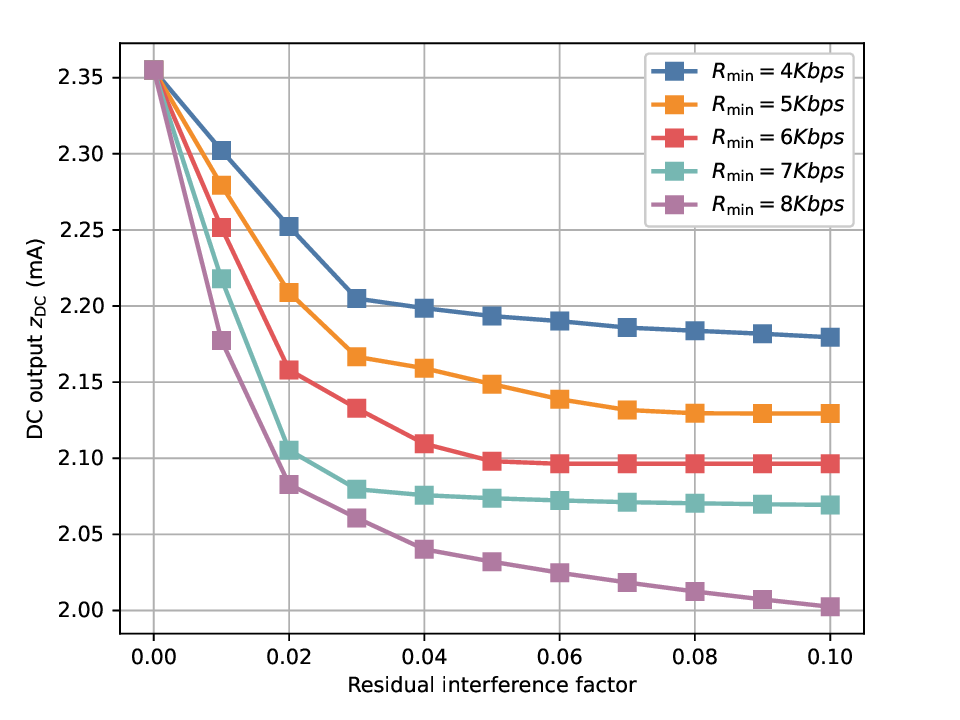}
			\vspace{-25pt}
			\caption{DC output versus residual interference factor $\lambda$ at $M=16$, $N=10$, $P_{\mathrm{tx}}=2~\mathrm{W}$, $d_{\mathrm{P}}=3~\mathrm{m}$, $d_{\mathrm{I}}=3~\mathrm{m}$ and $\Delta f=200~\mathrm{Hz}$.}
			\label{fig:lambda_R}
		\end{minipage}
	\end{figure*} 
	
		\begin{figure*}[htbp]
		\begin{minipage}{0.48\textwidth}
			\centering
			\includegraphics[width=\linewidth]{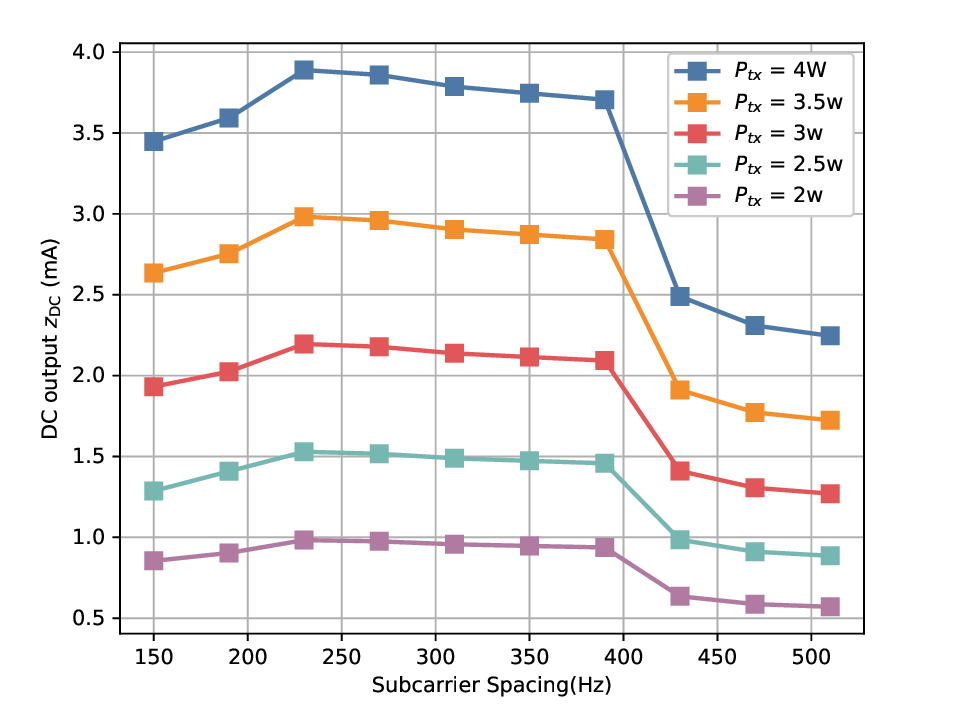}
			\vspace{-25pt}
			\caption{DC output versus subcarrier spacing $\Delta f$ at $M=10$, $N=10$, $d_{\mathrm{P}}=3~\mathrm{m}$, $d_{\mathrm{I}}=3~\mathrm{m}$, $R_{\mathrm{min}}=10000~\mathrm{bps}$ and $\lambda=0.02$. The number of 
				propagation paths $N_{\mathrm{path}}$ is set to 3.\protect\\ \phantom{Extra line}\protect\\ \phantom{Extra line}\protect\\ \phantom{Extra line}\protect\\ \phantom{Extra line}\protect\\ \phantom{Extra line}}
			\label{fig:Deltaf_P}
		\end{minipage}\hfill
		\begin{minipage}{0.48\textwidth}
			\centering
			\includegraphics[width=\linewidth]{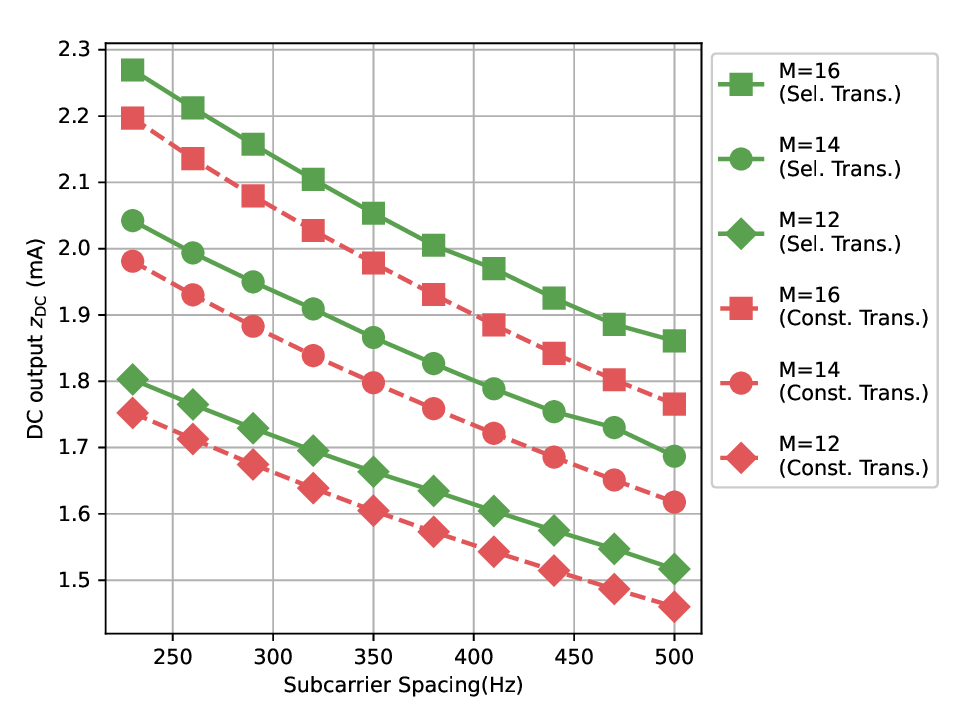}
			\vspace{-25pt}
			\caption{DC output versus subcarrier spacing $\Delta f$ at $N=10$, $P_{\mathrm{tx}}=2~\mathrm{W}$ $d_{\mathrm{I}}=3~\mathrm{m}$, $d_{\mathrm{P}}=3~\mathrm{m}$, $R_{\mathrm{min}}=1000~\mathrm{bps}$ and $\lambda=0.02$. The number of 
				propagation paths $N_{\mathrm{path}}$ is set to 3. The label ``Sel. Trans.'' denotes the case where the  frequency-selective response of the transducer is incorporated, whereas ``Const. Trans.'' denotes the case where the transducer conversion efficiency is assumed to be constant over all frequencies, as in \cite{xing2023performance}.}
			\label{fig:trans_compare}
		\end{minipage}
	\end{figure*} 
	
	\begin{figure*}[htbp]
		\begin{minipage}{0.48\textwidth}
			\centering
			\includegraphics[width=\linewidth]{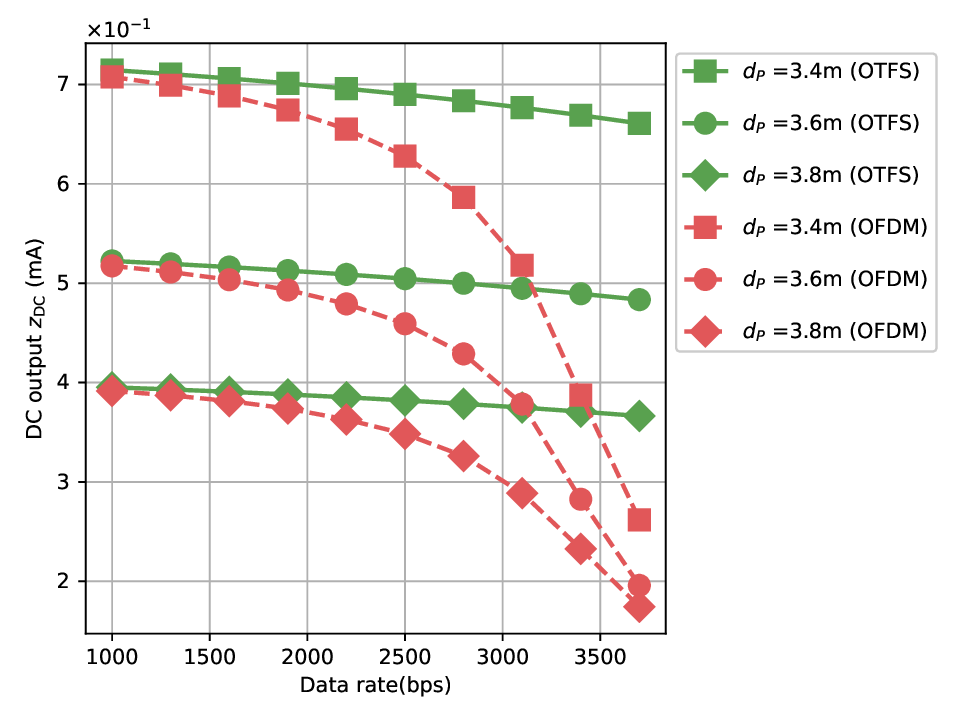}
			\vspace{-25pt}
			\caption{Performance comparison between OTFS and OFDM with $M=10$, $N=10$, $P_{\mathrm{tx}}=2~\mathrm{W}$, $d_{\mathrm{I}}=3~\mathrm{m}$, $\Delta f=200~\mathrm{Hz}$, and $\lambda=0.02$.\protect\\ \phantom{Extra line}}
			\label{fig:R_dP_OFDM}
		\end{minipage}\hfill
		\begin{minipage}{0.48\textwidth}
			\centering
			\includegraphics[width=\linewidth]{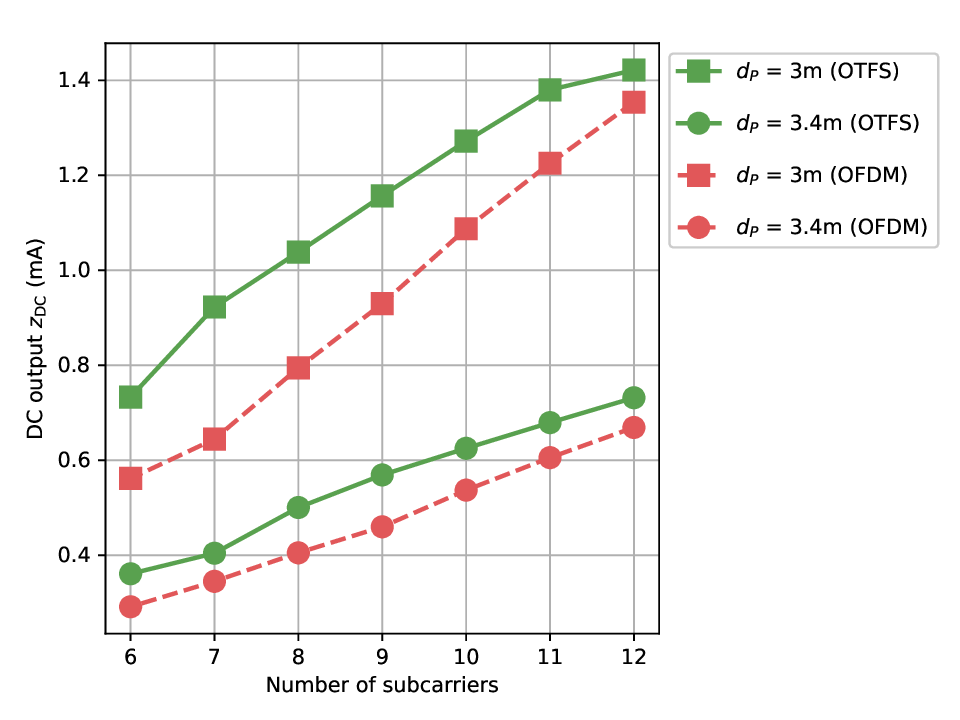}
			\vspace{-25pt}
			\caption{Performance comparison between OTFS and OFDM with $N=10$, $P_{\mathrm{tx}}=2~\mathrm{W}$, $d_{\mathrm{I}}=3~\mathrm{m}$, $\Delta f=400~\mathrm{Hz}$,  $R_{\min}=8000~\mathrm{bps}$ and $\lambda=0.04$. The number of 
				propagation paths $N_{\mathrm{path}}$ is set to 3.}
			\label{fig:M_dP_OFDM}
		\end{minipage}
	\end{figure*} 
	
	Fig.~\ref{fig:P_dP} shows the variation of the DC output $z_{\mathrm{DC}}$ with respect to the transmit power $P_{\mathrm{tx}}$ under different power-transfer distances $d_{\mathrm{P}}$. It can be observed that the DC output decreases as $d_{\mathrm{P}}$ increases. Meanwhile, for a fixed power-transfer distance, $z_{\mathrm{DC}}$ increases with the transmit power, and the growth exhibits a nonlinear trend. This is because a higher transmit power increases the received signal strength at the energy receiver, while the nonlinear rectifier can further exploit the higher-order components of the received waveform. Therefore, the DC output increases more significantly as the transmit power becomes larger. These results verify the nonlinear energy-harvesting characteristic of the rectifier and indicate that increasing the transmit power can effectively improve the acoustic power transfer performance of the proposed SAIPT system.
	
	Fig.~\ref{fig:lambda_R} shows the variation of the DC output $z_{\mathrm{DC}}$ with respect to the residual interference factor $\lambda$ under different data-rate requirements $R_\mathrm{min}$. It can be observed that the DC output decreases as $\lambda$ increases for all considered data-rate requirements. This is because a larger residual interference factor indicates that more uncancelled power-transfer interference remains at the information receiver, which increases the effective interference power in the achievable-rate expression. To satisfy the same communication requirement, the transmitter therefore needs to allocate more resources to information transfer, resulting in a reduction in the harvested DC output. In addition, a higher data-rate requirement leads to a lower DC output, since stricter communication constraints further limit the power and waveform degrees of freedom available for energy transfer. The decrease of $z_{\mathrm{DC}}$ is more pronounced when $\lambda$ is small, 
	because a slight increase in residual interference significantly increases the 
	effective interference power at the information receiver. As $\lambda$ becomes 
	larger, the waveform allocation is already strongly constrained by the rate 
	requirement, leaving limited room for further adjustment. Therefore, the 
	performance degradation gradually slows down, and the curves tend to flatten. These results demonstrate the importance of effective power-transfer waveform cancellation at the information receiver and reveal the tradeoff between interference suppression, information transfer, and acoustic power transfer in the proposed SAIPT system.

	Fig.~\ref{fig:Deltaf_P} shows the variation of the DC output $z_{\mathrm{DC}}$ with respect to the subcarrier spacing $\Delta f$ under different transmit power levels $P_{\mathrm{tx}}$. It can be observed that a larger transmit power always leads to a higher DC output, since more power can be exploited by the energy receiver for rectification. For a given transmit power, $z_{\mathrm{DC}}$ first increases with $\Delta f$ and then decreases when $\Delta f$ becomes large. This is because increasing $\Delta f$ enlarges the total communication bandwidth, which helps satisfy the data-rate requirement with less power allocated to information transfer, thereby leaving more power for energy transfer. However, when $\Delta f$ further increases, more subcarriers move away from the high-efficiency conversion band of the transducer, leading to reduced electro-acoustic and acoustic-electric conversion efficiencies. As a result, the harvested DC output decreases. This tradeoff indicates that an appropriate subcarrier spacing should be selected to balance the communication bandwidth gain and the transducer conversion efficiency in the proposed OTFS-based SAIPT system.
	
	Figs.~\ref{fig:trans_compare} illustrate the impact of the transducer frequency response on system performance under different numbers of subcarriers and subcarrier spacings. Due to the relatively low data-rate requirement in this setting, the power allocated to information-bearing symbols remains small, and thus its influence on the DC output is limited. As a result, the variation of system performance is mainly dominated by the energy transfer component rather than the information transmission component. In addition, the DC output decreases as the subcarrier spacing increases for all cases, since a larger spacing expands the occupied frequency band and pushes more subcarriers away from the transducer’s high-efficiency conversion region, thereby reducing the overall electro-acoustic and acousto-electric conversion efficiency. Moreover, the frequency-selective transducer model consistently achieves higher DC output than the constant-efficiency transducer model under identical parameter settings. The results confirms that accurately modeling the frequency-dependent characteristics of the transducer is essential for reliable performance evaluation and waveform design in SAIPT systems.
	
\begin{figure}[tbp]
		\centering
		\includegraphics[width=0.95\columnwidth]{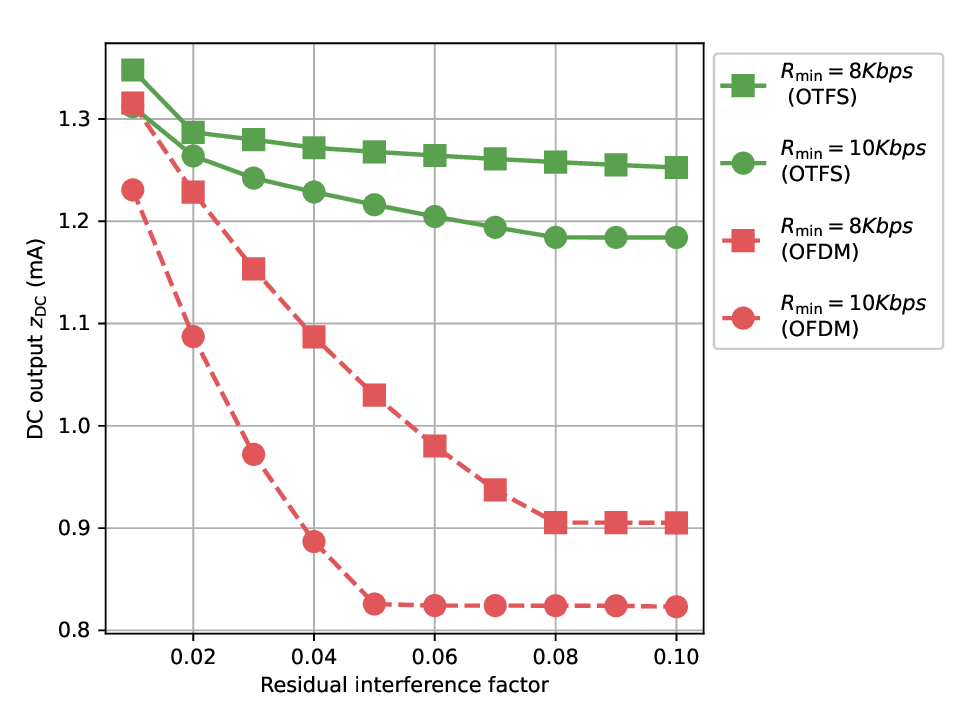}
		\caption{Performance comparison between OTFS and OFDM with $M=10$, $N=10$, $P_{\mathrm{tx}}=2~\mathrm{W}$, $d_{\mathrm{I}}=3~\mathrm{m}$, $d_{\mathrm{P}}=3~\mathrm{m}$,  and $\Delta f=400~\mathrm{Hz}$. The number of propagation paths $N_{\mathrm{path}}$ is set to 3.}
		\label{fig:lambda_R_OFDM}
	\end{figure}

	Figs.~\ref{fig:R_dP_OFDM}, \ref{fig:M_dP_OFDM} and~\ref{fig:lambda_R_OFDM} jointly compare the DC output performance of OTFS and OFDM under different system configurations. Fig.~\ref{fig:R_dP_OFDM} compares the DC output performance of OTFS and OFDM under different data-rate requirements and power-transfer distances. It can be observed that the DC output decreases as the power-transfer distance increases, which is mainly due to the more severe propagation loss over a longer underwater acoustic link. In addition, a higher data-rate requirement leads to a lower DC output for both schemes, since more transmit power needs to be allocated to the information-transfer symbols to satisfy the communication constraint, thereby reducing the power available for energy transfer. Compared with OFDM, OTFS consistently achieves a higher DC output under the same parameter settings. Moreover, as the required data rate increases, the DC output of OFDM decreases more rapidly, especially at larger data-rate requirements. This is because OFDM is more sensitive to the time-varying multipath underwater acoustic channel, and thus requires more power for reliable information transmission. In contrast, OTFS exploits the delay-Doppler domain representation and provides better robustness against channel variations, allowing more efficient waveform allocation between information transfer and power transfer. These results demonstrate the advantage of OTFS-based SAIPT waveform design in dynamic underwater acoustic channels.
	
	Fig.~\ref{fig:M_dP_OFDM} depicts the DC output performance versus the number of subcarriers for  OTFS and OFDM schemes under different power-transfer distances. It can be observed that the DC output increases with the number of subcarriers in all cases. This is because a larger number of subcarriers generally results in a higher PAPR of the multicarrier signal, which strengthens the nonlinear rectification effect and improves the DC output. Therefore, the multicarrier waveform can effectively exploit the nonlinear characteristic of the rectifier, and the energy-harvesting performance can be enhanced by increasing the number of subcarriers. In addition, increasing the power-transfer distance from 3 m to 3.4 m results in a clear degradation of DC output for both schemes, which is mainly attributed to the increased propagation loss in underwater acoustic channels. Furthermore, the OTFS-based scheme consistently achieves higher DC output than the OFDM-based scheme. The performance gain is more pronounced when the number of subcarriers is small, since OFDM requires more transmit power to ensure reliable communication under time-varying underwater channels, leaving less power available for energy harvesting. In contrast, OTFS provides more robust transmission in the delay–Doppler domain, enabling better energy–information tradeoff.

	Fig.~\ref{fig:lambda_R_OFDM} illustrates the DC output versus residual interference factor $\lambda$ under different data-rate requirements $R_\mathrm{min}$ for both OTFS and OFDM schemes. It can be observed that a higher data-rate requirement results in a lower DC output for both schemes due to the increased communication power demand. In addition, the DC output decreases with increasing residual interference factor, as stronger interference degrades the achievable rate performance and forces more transmit power to be allocated to information-transfer symbols, thereby reducing the power available for energy harvesting. Moreover, the performance gap between OTFS and OFDM becomes more pronounced as $\lambda$ increases, since OTFS is more robust to residual interference in the delay–Doppler domain, whereas OFDM suffers more severe interference spreading in the time–frequency domain, leading to faster degradation in energy harvesting performance. In addition, the DC output of OFDM becomes nearly saturated when $\lambda$ is large, because almost all transmit power is consumed to satisfy the minimum data-rate requirement under strong interference conditions. The results in these figures consistently demonstrate the superiority of OTFS over OFDM in underwater acoustic channels characterized by severe multipath propagation and Doppler effects.

\section{CONCLUSION} \label{sec:CONCLUSION}
	In this paper, an OTFS-assisted SAIPT waveform design was investigated for dynamic underwater acoustic channels by exploiting the delay-Doppler domain representation to enhance robustness against time-varying multipath propagation and Doppler effects. The electro-acoustic and acoustic-electric conversion efficiencies of the transducer were modeled, and a nonlinear rectifier model was incorporated to characterize the harvested DC output. Based on the derived achievable-rate and DC-output expressions, a joint waveform optimization problem was formulated to maximize the DC output under the transmit power and minimum data-rate constraints. To address the non-convexity of the problem, an SCA-based iterative algorithm was developed. Simulation results showed that the proposed OTFS-based SAIPT design achieves higher DC output than the OFDM-based scheme, especially under stringent data-rate requirements. In addition, the effects of key system parameters, including the number of subcarriers, transmit power, power-transfer distance, residual interference factor, and subcarrier spacing, were analyzed, demonstrating the importance of jointly considering the transducer frequency response, rectifier nonlinearity, and multicarrier waveform characteristics in underwater acoustic power transfer.

\bibliographystyle{IEEEtran}
\bibliography{IEEEabrv,reference}
\end{document}